\documentclass{aa}
\usepackage{graphicx}
\usepackage{amssymb}
\usepackage{amsmath}
\usepackage{natbib}
\usepackage{mathrsfs}
\usepackage{ulem}
\usepackage{txfonts}
\usepackage{lmodern}
\usepackage{booktabs}

\usepackage[switch]{lineno} 

\newcommand{\cible}{{KIC7581399}}
\newcommand{\kepler}{\textit{Kepler}}

\newcommand{\ind}[1]{_{\rm #1}}

\def\m2s2{\,m$^{2}$\,s$^{-2}$} 
\def\kms{\,km\,s$^{-1}$}       

\newcommand{\vaisala}{Brunt-V\"ais\"al\"a}

\newcommand{\losc}{\textsc{losc}}

\newenvironment{itemize*}%
  {\begin{itemize}%
    \setlength{\itemsep}{1pt}%
    \setlength{\parskip}{1pt}}%
  {\end{itemize}}

\newcommand\T{\rule{0pt}{2.6ex}}
\newcommand\B{\rule[-1.2ex]{0pt}{0pt}}
\newcommand\TT{\rule{0pt}{4ex}}
\newcommand\BB{\rule[-2.8ex]{0pt}{0pt}}

\begin{document}
\title{Seismic evidence for a weak radial differential rotation in intermediate-mass core helium burning  stars}
\titlerunning{Differential rotation in \textit{Kepler} secondary clump stars}
\author{
S. Deheuvels\inst{1,2}
\and J. Ballot\inst{1,2}
\and P.~G. Beck\inst{3}
\and B. Mosser\inst{4}
\and R. \O stensen\inst{5}
\and R.~A. Garc\'{\i}a\inst{3}
\and M.~J. Goupil\inst{4}
}

\institute{Universit\'e de Toulouse; UPS-OMP; IRAP; Toulouse, France
\and CNRS; IRAP; 14, avenue Edouard Belin, F-31400 Toulouse, France
\and Laboratoire AIM Paris-Saclay, CEA/DSM-CNRS-Universit\'e Paris Diderot, IRFU/SAp, Centre de Saclay, 91191 Gif-sur-Yvette cedex, France 
\and LESIA, Observatoire de Paris, PSL Research University, CNRS, Universit\'e Pierre et Marie Curie, Universit\'e Denis Diderot,  92195 Meudon cedex, France 
\and Instituut voor Sterrenkunde, KU Leuven, 3001 Leuven, Belgium 
}

\offprints{S. Deheuvels\\ \email{sebastien.deheuvels@irap.omp.eu}
}


\abstract{The detection of mixed modes that are split by rotation in \kepler\ red giants has made it possible to probe the internal rotation profiles of these stars, which brings new constraints on the transport of angular momentum in stars. Mosser et al. (2012) have measured the rotation rates in the central regions of intermediate-mass core helium burning stars (secondary clump stars).}
{Our aim is to exploit the rotational splittings of mixed modes to estimate the amount of radial differential rotation in the interior of secondary clump stars using \kepler\ data, in order to place constraints on angular momentum transport in intermediate-mass stars.}
{We select a subsample of \kepler\ secondary clump stars with mixed modes that are clearly rotationally split. By applying a thorough statistical analysis, we show that the splittings of both gravity-dominated modes (trapped in central regions) and p-dominated modes (trapped in the envelope) can be measured. We then use these splittings to estimate the amount of differential rotation by using inversion techniques and by applying a simplified approach based on asymptotic theory (Goupil et al. 2013).}
{We obtain evidence for a weak radial differential rotation for six of the seven targets that were selected, with the central regions rotating $1.8\pm0.3$ to $3.2\pm1.0$ times faster than the envelope. The last target is found to be consistent with a solid-body rotation.}
{This demonstrates that an efficient redistribution of angular momentum occurs after the end of the main sequence in the interior of intermediate-mass stars, either during the short-lived subgiant phase, or once He-burning has started in the core. In either case, this should bring constraints on the angular momentum transport mechanisms that are at work.}


\maketitle

\section{Introduction}

Even though rotation is known to have an important impact on stellar structure and evolution, we still lack a theoretical understanding of how the internal rotation rates of stars evolve in time. Several physical processes can transport angular momentum (AM) in stars: hydrodynamical instabilities and meridional circulation (\citealt{zahn92}, \citealt{mathis04}), internal gravity waves (IGW) excited at the edge of convective regions (\citealt{charbonnel05}), and magnetic mechanisms, either from a buried fossil magnetic field (\citealt{gough98}) or from magnetic instabilities (\citealt{spruit99}, \citealt{rudiger14a}). It is now well known that rotationally-induced mechanisms of AM as they are currently understood are far too inefficient to produce a rigid rotation for the solar radiative interior as found by helioseismology (\citealt{schou98}, \citealt{chaplin99}, \citealt{garcia08}).

The asteroseismic measurement of the internal rotation profiles of red giants has recently shed some new light on this long-standing problem. The very high precision reached by the \kepler\ space mission has made it possible to measure the rotational splittings of mixed modes, i.e. of modes that behave both as pressure modes in the envelope and as gravity modes in the core. It was thus shown that stellar cores spin up during the subgiant phase (\citealt{deheuvels14}, hereafter D14) and that a significant radial differential rotation establishes by the time stars reach the red giant branch (RGB) (\citealt{beck12}, \citealt{deheuvels12}, hereafter D12, \citealt{beck14}). This was expected as a consequence of the severe core contraction that occurs after the main-sequence turnoff. However, the core rotation rates measured for red giants are much slower than they would be in the absence of AM transport. Besides, by measuring the core rotation of several hundreds of red giants, \cite{mosser12b} showed that even though stellar cores keep contracting, they actually spin down during the evolution along the RGB. These observations clearly indicate that a very efficient mechanism of AM transport operates between the core and the envelope of red giants. 

For now it seems that none of the known mechanisms of AM transport are able to account on their own for the evolution of the rotation profiles of low-mass stars in the RGB obtained from seismology. Purely hydrodynamical processes including the effects of meridional circulation and turbulence shear as currently understood predict core rotation rates that are 2 to 3 orders of magnitude higher than the observations (\citealt{eggenberger12}, \citealt{marques13}, \citealt{ceillier13}). Magnetic instabilities were also considered as a possible source of AM transport. The instability of toroidal magnetic fields under the influence of differential rotation (azimuthal magnetorotational instability, AMRI) is a promising candidate. \citet{rudiger14b,rudiger14a} showed that the resulting effective viscosity may be high enough to account for the slow core rotation rate of red giants, but stratification needs to be taken into account in order to quantify its effect. On the other hand, \cite{cantiello14} showed that even if the so-called Tayler-Spruit (\citealt{spruit02}) dynamo loop exists, which is still debated, it does not generate sufficient AM transport to reproduce the slow core rotation rates observed in \kepler\ red giants. Interestingly, the transport of AM through IGW excited at the bottom of the convective envelope could account for the core/envelope decoupling during the subgiant phase. During this phase, IGW are expected to damp out and release their AM before they reach the stellar core owing to the large increase in the \vaisala\ frequency, which could decouple the core from the envelope (\citealt{talon08}, \citealt{fuller14}). However, IGW seem unable to then produce the recoupling that is necessary to account for the spin down of the core observed by \cite{mosser12b}. 

So far, most of the constraints obtained from the seismology of red giants involve low-mass stars ($M\lesssim 2.4 M_\odot$) which ascend the RGB before triggering He-burning in a flash due to the degeneracy of the core. These stars are known as \textit{red clump} stars in reference to their location in the HR diagram. Less constraints exist on the internal rotation of intermediate-mass stars. Recently, \cite{kurtz14} were able to seismically measure the internal rotation of a main sequence A-type star and found that this star rotates almost rigidly. However, it should be noted that this star might not be representative of all A-type stars since it has a peculiarly slow rotation rate. After the end of the MS, intermediate-mass stars cross the subgiant phase and the bottom of the RGB on a Kelvin-Helmoltz timescale, before settling in the clump (they are then referred to as \textit{secondary clump} stars). This raises the question whether AM transport can be fast enough to operate on such a short timescale. If not these stars should reach the clump with strong differential rotation and very fast-spinning cores. \cite{mosser12b} were able to seismically measure the mean rotation rate in the g-mode cavity for secondary clump stars and found periods ranging from 20 to 250 days. One must keep in mind that the He-burning core is convective, so gravity waves do not propagate inside the core and the rotation rates obtained by \cite{mosser12b} correspond to the layers just above the He-burning convective core. Assuming solid-body (SB) rotation at the end of the main sequence, \cite{tayar13} showed that the rotation rates found by \cite{mosser12b} are compatible with a SB-rotation profile in the clump for intermediate-mass stars, which suggests that an efficient AM redistribution has operated by the time these stars reach the clump.

In this paper, we aim at testing the hypothesis of a SB-rotation in the interior of intermediate-mass clump stars with seismology, using \kepler\ data. It has been shown that a ratio between the mean core rotation and the mean envelope rotation can be estimated from the splittings of mixed modes in subgiants and young red giants, see \cite{goupil13} (hereafter G13), D14. On the other hand, for more evolved red giants the rotational splittings are dominated by the contribution of the central layers, and no reliable estimate of the envelope rotation can be obtained (G13). We here show that constraints on the amount of differential rotation can be obtained for clump stars, provided the splittings of p-dominated modes can be reliably estimated. This latter condition is not obviously satisfied for clump stars because the mode widths of their p-dominated modes are comparable to the rotational splittings. We thus dedicate special care to establishing the significance of our estimates of the splittings of these modes. In Sect. \ref{sect_targets}, we select the red-clump stars that are the most likely to provide constraints on the amount of differential rotation among the \kepler\ targets. We then proceed to estimate the mode frequencies and rotational splittings in Sect. \ref{sect_analysis}, and we address the question of the significance of the estimated splittings. In Sect. \ref{sect_interpretation}, we validate the simplified approach proposed by G13 to measure the internal rotation using the observed splittings and we apply it to all the stars of our sample. We thus find evidence for a weak radial differential rotation in six of the seven selected targets, the last one being consistent with a SB rotation. In Sect. \ref{sect_spectro} we try to obtain complementary information on the rotation profiles of three stars of the sample based on high-resolution spectroscopic observations (HERMES spectrograph), but the constraints we obtain are too loose to test the seismic results. We finally discuss the implications of our results in terms of AM transport in intermediate-mass stars in Sect. \ref{sect_conclusion}.

\section{Selection and characterization of targets \label{sect_targets} \label{sect_select}}

Among the red giants observed with \kepler, we searched for secondary clump stars that are most likely to provide constraints on the amount of differential rotation. For this purpose, we used the complete \kepler\ dataset available, i.e. 4 years of data from quarters Q0 through Q17 with the long cadence mode (integration time of 29.4 min). The light curves were processed using the \kepler\ pipeline developed by \cite{jenkins10}, and corrections from outliers, occasional jumps, and drifts were also applied (\citealt{garcia11}). The power density spectra were obtained by using the Lomb-Scargle periodogram (\citealt{lomb76}, \citealt{scargle82}), and we then applied the screening process that is described in the sections below in order to identify optimal targets.

\subsection{Selection based on stellar mass}

We first applied a mass criterion to select secondary clump stars. Standard stellar models predict that stars that form with a mass over 2.4 $M_\odot$ trigger He-burning in a non-degenerate core. However, non-standard physical processes such as core overshooting during the MS can have a large influence on this theoretical threshold (\citealt{montalban13}). Besides, stars are expected to lose mass during their ascent of the RGB, so that the lower limit on the masses of secondary clump stars is in fact lower than the predicted theoretical threshold. Based on the properties of the observed mixed modes in \kepler\ secondary clump giants, \cite{mosser14} observationally estimated the mass-limit to about $1.9\pm0.2\,M_\odot$ (see Fig. 1 of their paper). In this study, we used the upper bound of this observational threshold in order to ensure that only secondary clump stars are retained, and selected stars with masses above 2.1 $M_\odot$. For this purpose, we applied seismic scaling relations, which required us to estimate the mean large separation of acoustic modes $\Delta\nu$ and the frequency of maximum power of solar-like oscillations $\nu_{\rm max}$ in the considered targets. An estimate of the effective temperature is also needed, which we obtain photometrically following \cite{pinsnneault12} (see Sect. \ref{sect_scalinglaws}).

\subsubsection{Estimate of $\nu_{\rm max}$ \label{sect_numax}}


In order to estimate the frequency of maximum power of the oscillations, we fitted a Gaussian envelope to the power spectrum in the neighborhood of the detected solar-like oscillations using a maximum likelihood estimate (MLE) method. The central frequency of the fitted Gaussian function provides an estimate of $\nu_{\rm max}$. The fit required to include a description of the background. The contribution from granulation was modeled as a Harvey profile (\citealt{harvey85}), and white noise was added to account for photon noise. As was pointed out by \cite{karoff13} and \cite{kallinger14}, we found that an additional Harvey profile was required to correctly fit the observed background. Beside providing an estimate of $\nu_{\rm max}$, this fit yields a model for the background, which we further use to extract the properties of the oscillation modes (Sect. \ref{sect_analysis}).

\subsubsection{Estimate of $\Delta\nu$ \label{sect_deltanu}}

To estimate the mean large separation $\Delta\nu$ of the targets, we followed the procedure of \cite{mosser13}. It consists in fitting a second-order asymptotic expression to the observed modes. Instead of including the second-order terms as prescribed by \cite{tassoul80}, the authors proposed to introduce a constant parameter $\alpha$ corresponding to a linear variation of the large separation with radial order $n$. This yields the following expression for radial modes
\begin{linenomath*}
\begin{equation}
\nu_{n,l=0} = \left[ n + \varepsilon_{\rm p} + \frac{\alpha}{2}(n-n_{\rm max})^2 \right] \Delta\nu 
\label{eq_purep}
\end{equation}
\end{linenomath*}
The index $n_{\rm max}$ was chosen so that it matches the frequency of maximum power of the oscillations $\nu_{\rm max}$, and it is therefore not necessarily an integer.
By smoothing the power spectrum of the observed stars, we obtained first approximate estimates of the frequencies of $l=0$ modes.
The expression given by Eq. \ref{eq_purep} was then fitted to our estimates of the $l=0$ mode frequencies, yielding optimal values of $\Delta\nu$, $\varepsilon_{\rm p}$, and $\alpha$. An example of this fit is shown in Fig. \ref{fig_echelle_obs} for one star of the sample. This provides a precise observational value of $\Delta\nu$, which can be translated in the asymptotic value if needed.

\subsubsection{Seismic scaling relations \label{sect_scalinglaws}}

We then used seismic scaling relations that relate the global seismic parameters $\Delta\nu$ and $\nu_{\ind{max}}$ to stellar properties such as the mass, radius and surface gravity (\citealt{brown91}). These relations are derived from the fact that $\nu_{\rm max}$ scales to good approximation as the acoustic cut-off frequency (\citealt{brown91}, \citealt{stello08}, \citealt{belkacem11}). To obtain estimates of the stellar masses and radii of the considered stars, an estimate of the effective temperature of the star was also needed. We have used the photometric estimates of $T_{\rm eff}$ obtained from the recipe proposed by \cite{pinsonneault12} that we applied to the \textit{griz} photometry in the \kepler\ input catalog (KIC). As mentioned above, we retained only the stars that have a mass above 2.1 $M_\odot$.

\subsection{Selection based on the evolutionary stage \label{sect_deltapi}}


\begin{figure}
\begin{center}
\includegraphics[width=9cm]{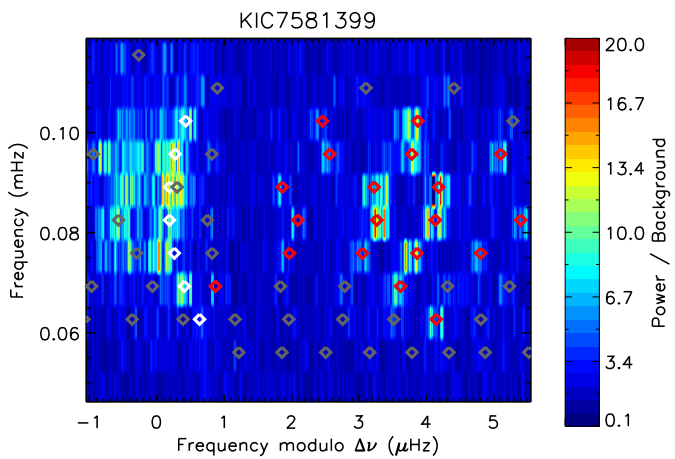}
\end{center}
\caption{\'Echelle diagram of KIC7581399, obtained from 4 years of \kepler\ data. The white diamonds correspond to a fit of the asymptotic expression of radial modes given by Eq. \ref{eq_purep} to the observed radial modes. The gray and red diamonds correspond to a fit of the asymptotic expression of mixed modes given by Eq. \ref{eq_matching}, \ref{eq_thetap_m12} and \ref{eq_thetag_m12} (only the red modes were used to perform the fit).  
\label{fig_echelle_obs}}
\end{figure}

To select secondary clump stars, we needed to distinguish core-He-burning (clump) stars from H-shell-burning (RGB) stars. As is well known, this cannot be achieved merely from the location of stars in the HR diagram. However, these two classes of stars have a very different core structure (clump stars have a convective core due to core-He burning). As a result, the mean period spacing of g modes for these two classes of stars are sufficiently different to distinguish them spectacularly well (\citealt{bedding11}). We thus evaluated the period spacing $\Delta\Pi_1$ of $l=1$ modes in the considered stars. For this purpose, we followed the recipe that \cite{mosser12a} adapted from the asymptotic expression of p-g mixed modes of \cite{unno89}. According to Eq. 16.50 of \cite{unno89}, the matching in the evanescent zone of solutions corresponding to g modes in the core and to p modes in the envelope requires that
\begin{linenomath*}
\begin{equation}
\tan(\theta_{\rm p}) = q \tan(\theta_{\rm g})
\label{eq_matching}
\end{equation}
\end{linenomath*}
where
\begin{linenomath*}
\begin{equation}
\theta_{\rm g} \equiv \int_{r_{\rm a}}^{r_{\rm b}} k_r \,\hbox{d}r  , \; \hbox{and} \; \theta_{\rm p} \equiv \int_{r_{\rm c}}^{r_{\rm d}} k_r \,\hbox{d}r
\label{eq_thetapg}
\end{equation}
\end{linenomath*}
and the radii $r_{\rm a}$ and $r_{\rm b}$ (resp. $r_{\rm c}$ and $r_{\rm d}$) are the inner and outer turning points of the g-mode (resp. p-mode) cavity. The parameter $q$ corresponds to the coupling between the p- and g-mode cavities. Its variations with frequency are neglected. \cite{mosser12a} have developed the expressions of the phases $\theta_{\rm p}$ and $\theta_{\rm g}$ in order to take into account higher-order terms in the asymptotic developments of p and g modes, yielding
\begin{linenomath*}
\begin{align}
\theta_{\rm p} & = \frac{\pi}{\Delta\nu} \left[ \nu - \left( \nu_{\rm p} \right)_{n,1} \right] \label{eq_thetap_m12} \\
\theta_{\rm g} & = \pi \left( \frac{1}{\Delta\Pi_1\nu} - \varepsilon_{\rm g} \right) \label{eq_thetag_m12} 
\end{align}
\end{linenomath*}
where the $\left(\nu_{\rm p} \right)_{n,1}$ correspond to the frequencies of theoretical pure $l=1$ p modes. The values of $\left(\nu_{\rm p} \right)_{n,1}$ can be obtained from the second-order asymptotic expression of radial p modes as follows
\begin{linenomath*}
\begin{equation}
\left(\nu_{\rm p} \right)_{n,1} = \nu_{n,l=0} +(1/2 - d_{01}) \Delta\nu
\end{equation}
\end{linenomath*}
where $d_{01}$ corresponds to the mean small separation built with $l=0$ and $l=1$ modes. The value of this quantity is a priori not known since pure $l=1$ p modes cannot be observed.

\begin{table}
  \begin{center}
  \caption{Period spacings \label{tab_deltapi}}
\begin{tabular}{l c c c c}
\hline \hline
\T \B KIC-Id & $\Delta\Pi_1$ (s) & $q$ & $\varepsilon_{\rm g}$ & $d_{01}$ ($\mu$Hz) \\
\hline 
\T KIC8962923 & 299.8 & 0.27 & 0.76 & 0.34 \\
KIC5184199 & 239.7 & 0.22 & 0.91 & 0.48 \\
KIC4659821 & 214.6 & 0.18 & 0.66 & 0.51 \\
KIC3744681 & 233.1 & 0.41 & 0.45 & 0.16  \\
KIC7467630 & 293.5 & 0.19 & 0.85 & 0.39 \\
KIC9346602 & 241.4 & 0.23 & 0.43 & 0.47 \\
\B KIC7581399 & 222.4 & 0.18 & 0.74 & 0.47 \\
\hline
\end{tabular}
\end{center}
\end{table}

For a given period spacing $\Delta\Pi_1$, coupling term $q$, phase term $\varepsilon_{\rm g}$, and $d_{01}$ parameter, the mode frequencies can be approximated by the solutions of Eq. \ref{eq_matching} that can be obtained with a Newton-Raphson algorithm. We then searched for the combination ($\Delta\Pi_1, q, \varepsilon_{\rm g},d_{01}$) that reproduces the observed modes at best. The results are given for each star in Table \ref{tab_deltapi}. The mode frequencies obtained with the optimal set of parameters for one target of the sample are shown in Fig. \ref{fig_echelle_obs}. We note that we have further used our estimates of $\Delta\Pi_1$, $q$, and $\varepsilon_{\rm g}$ to evaluate the trapping of the modes in Sect. \ref{sect_G13}.

We retained only the stars for which the mean period spacing $\Delta\Pi_1$ was found to be above 120 s in order to select only clump stars (see \citealt{bedding11}, \citealt{mosser14}).


\subsection{Selection based on the rotational splittings of the modes}

\begin{table*}
  \begin{center}
  \caption{Global parameters of selected targets \label{tab_targets}}
\begin{tabular}{l c c c c c c}
\hline \hline
\T \B KIC-Id & $\Delta\nu$ ($\mu$Hz) & $\nu_{\rm max}$ & $T_{\rm eff}$ (K) & $M/M_\odot$ & $R/R_\odot$ & $\log g$ \\
\hline 
\T KIC8962923 & $6.95 \pm0.13$ & $ 79.3\pm  0.3$  & $5253\pm 91$ & $2.17\pm0.25$ & $ 9.36\pm0.48$ & $2.832\pm0.013$ \\
KIC5184199 & $7.85\pm0.13$ & $94.7\pm0.4$  & $5197\pm89$ & $2.23\pm0.23$ & $8.70\pm0.40$ & $2.907\pm0.012$ \\
KIC4659821 & $8.19\pm0.09$ & $101.5\pm0.3$  & $5160\pm98$ & $2.29\pm0.19$ & $8.53\pm0.30$ & $2.935\pm0.013$ \\
KIC3744681 & $5.47\pm0.13$ & $61.1\pm0.5$  & $5084\pm68$ & $2.46\pm0.35$ & $11.45\pm0.73$ & $2.712\pm0.015$ \\
KIC7467630 & $6.06\pm0.09$ & $70.6\pm0.4$  & $5114\pm103$ & $2.55\pm0.27$ & $10.81\pm0.49$ & $2.776\pm0.015$ \\
KIC9346602 & $5.07\pm0.13$ & $55.7\pm0.3$  & $5162\pm79$ & $2.58\pm0.37$ & $12.22\pm0.80$ & $2.675\pm0.013$ \\
\B KIC7581399 & $6.57\pm0.13$ & $80.9\pm0.4$  & $5301\pm80$ & $2.92\pm0.35$ & $10.73\pm0.57$ & $2.843\pm0.013$ \\
\hline
\end{tabular}
\end{center}
\end{table*}

Finally, an obvious condition to measure the amount of differential rotation was that the oscillation modes of the stars should be clearly split by rotation so that reliable rotational splittings can be extracted from the power spectra. A first inspection of \kepler\ oscillation spectra immediately shows that such stars are harder to find among clump stars than among RGB stars. There are two reasons for this: (1) the lifetimes of the modes are shorter for clump stars, so that they have larger widths, and (2) as shown by \cite{mosser12a}, the cores of clump stars have longer rotation periods than those of RGB stars, which means that their rotational splittings are smaller. As a result of these two combined effects, the rotational splittings of the modes are often comparable to their widths, which complicates  the measurement of splittings. This is not so much a problem for g-dominated modes, which have larger inertias and therefore shorter widths. However, the situation is more complicated for p-dominated modes. We stress that the splittings of these latter modes are absolutely necessary to estimate the amount of differential rotation between the core and the envelope. We therefore performed a first pre-selection of targets based on a visual inspection of the \'echelle diagrams of all the considered giants to identify those whose mixed modes are most clearly split. We note that the most favorable stars are those that are seen equator-on (inclination angle $i=90^\circ$), because in this case only the $m=\pm1$ components of $l=1$ modes, which are separated by twice the rotational splitting, are visible.

The characteristics of the selected targets are listed in Table \ref{tab_targets}. Only the targets that provided enough reliable splittings (see Sect. \ref{sect_analysis}) to estimate differential rotation are mentioned. The following study is based on this sample of seven targets.

\section{Extraction of rotational splittings \label{sect_analysis}}

In order to measure the amount of differential rotation in the selected targets, we needed to extract the rotational splittings of the modes. For this purpose, we used a maximum likelihood estimate (MLE) method in the same way as D14. We here only briefly summarize it. We performed individual fits of each rotational multiplet (modes of same radial order $n$ and degree $l$). Among the multiplets, the $m$-components were modeled as Lorentzian profiles. The components were assumed to have a common width $\Gamma$. Within the multiplets, the height ratios between the components are given by a visibility factor that depends on the inclination angle $i$ of the star (\citealt{gizon03}, \citealt{ballot06}). Finally, we assumed that the multiplets are symmetric with respect to the central $m=0$ component. They are thus equally spaced by the rotational splitting whose expression is given by
\begin{linenomath*}
\begin{equation}
\left(\delta\nu_{\rm s}\right)_{n,l} = \int_0^R K_{n,l}(r) \frac{\Omega(r)}{2\pi} \,\hbox{d}r
\label{eq_splitting}
\end{equation}
\end{linenomath*}
where the functions $K_{n,l}(r)$ are the rotational kernels, which depend on the eigenfunctions of the modes.

As was mentioned in Sect. \ref{sect_select}, one of the main challenges of this work was to establish that the splittings of p-dominated modes can be recovered for clump stars. For the selected targets, several p-dominated modes are visually split (see Fig. \ref{fig_pmodes})\footnote{The trapping of the modes is evaluated in Sect. \ref{sect_interpretation}}. However, owing to the larger linewidths of these modes, it is not clear at first sight whether this is caused by the rotational splitting of the modes, or only the result of stochastic excitation. To distinguish between these two cases, we performed statistical tests, following both a frequentist and a Bayesian approach in a complementary way.

\subsection{Frequentist approach \label{sect_MLE}}

The frequentist approach consists in computing a false-alarm probability using the so-called \textit{$H_0$ test}. For each of the observed $l=1$ multiplets in the oscillation spectra of the selected targets, we performed two MLE fits using successively the two following hypotheses:


\begin{figure*}
\begin{center}
\includegraphics[width=7cm]{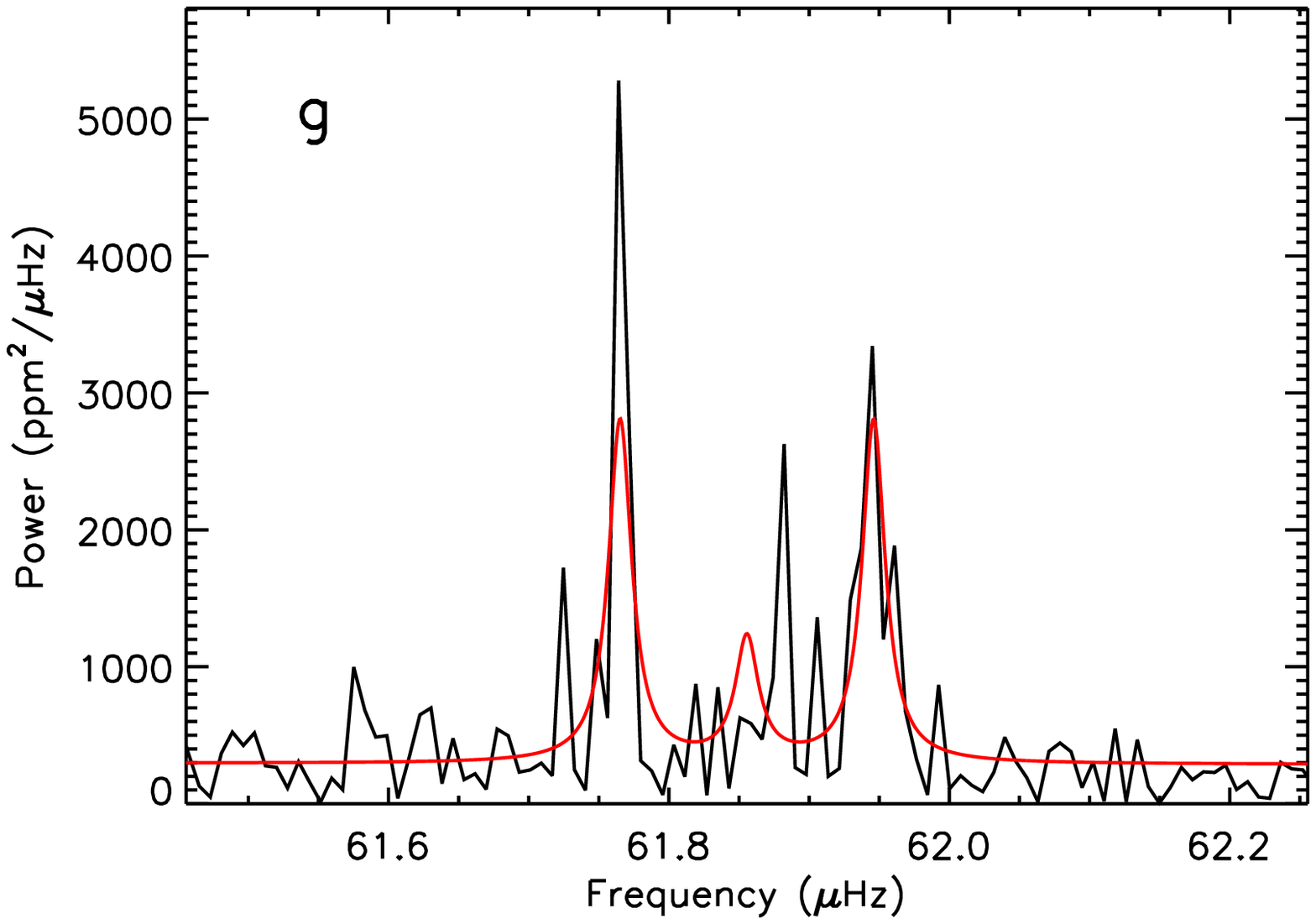}
\includegraphics[width=7cm]{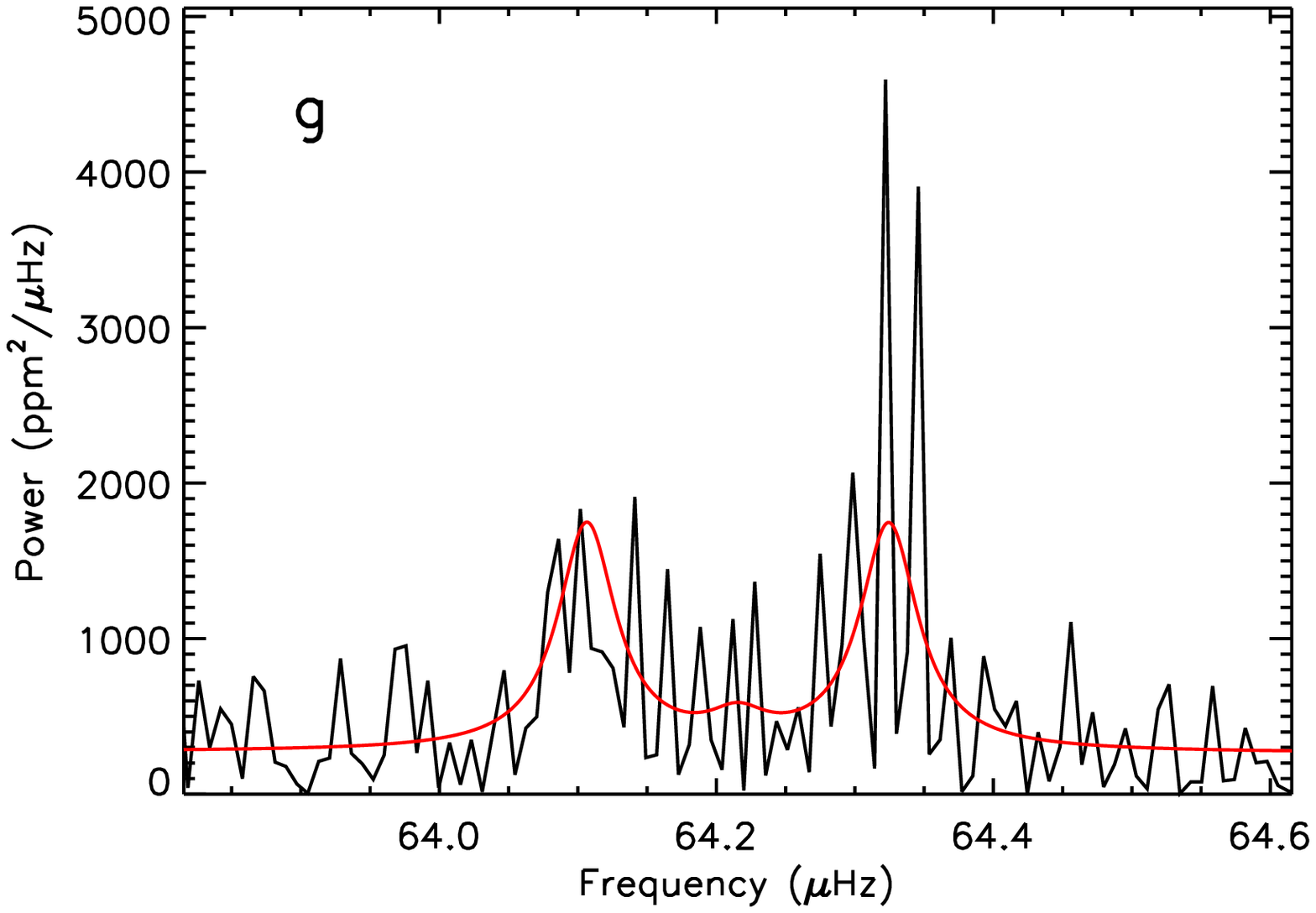}
\includegraphics[width=7cm]{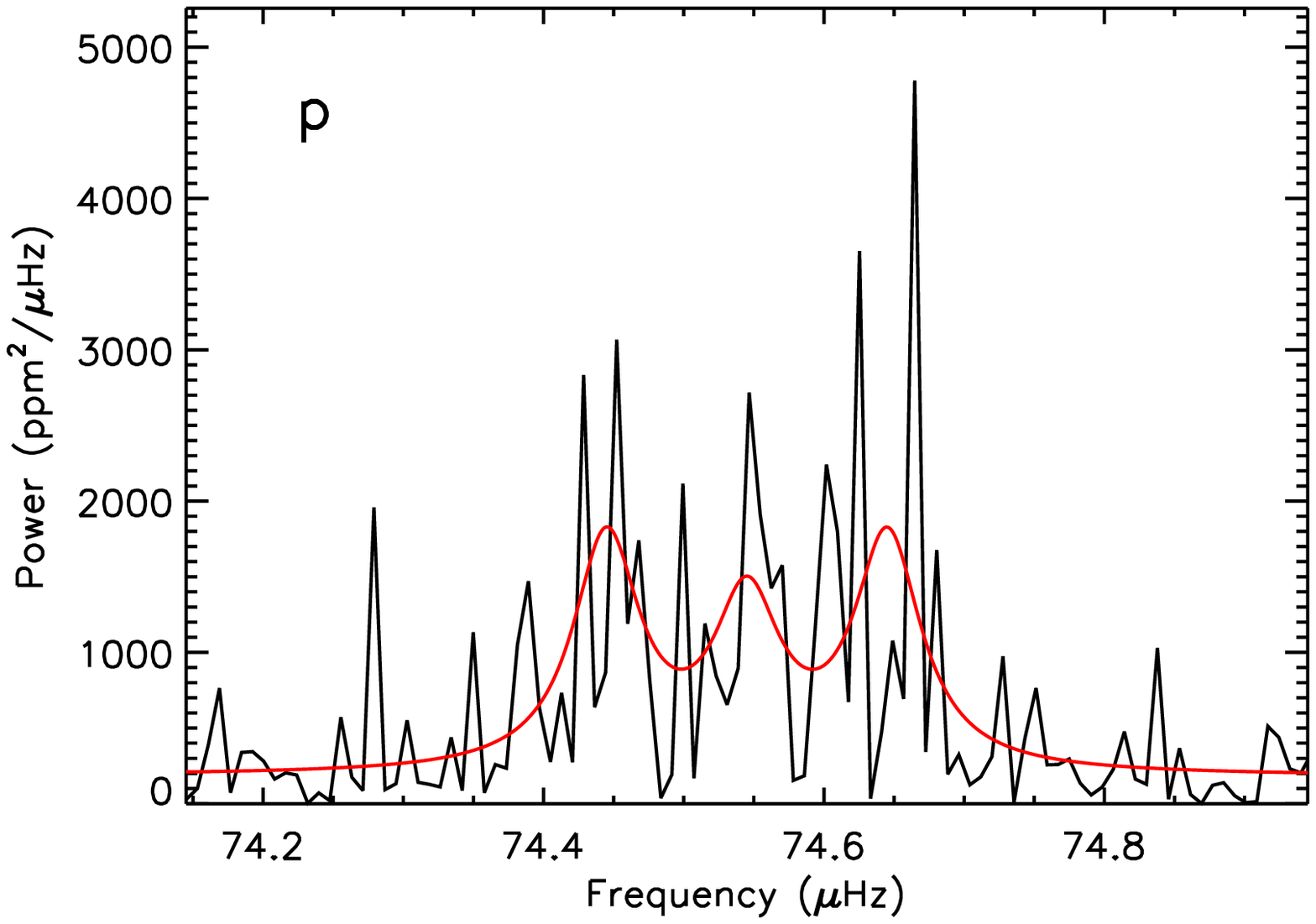}
\includegraphics[width=7cm]{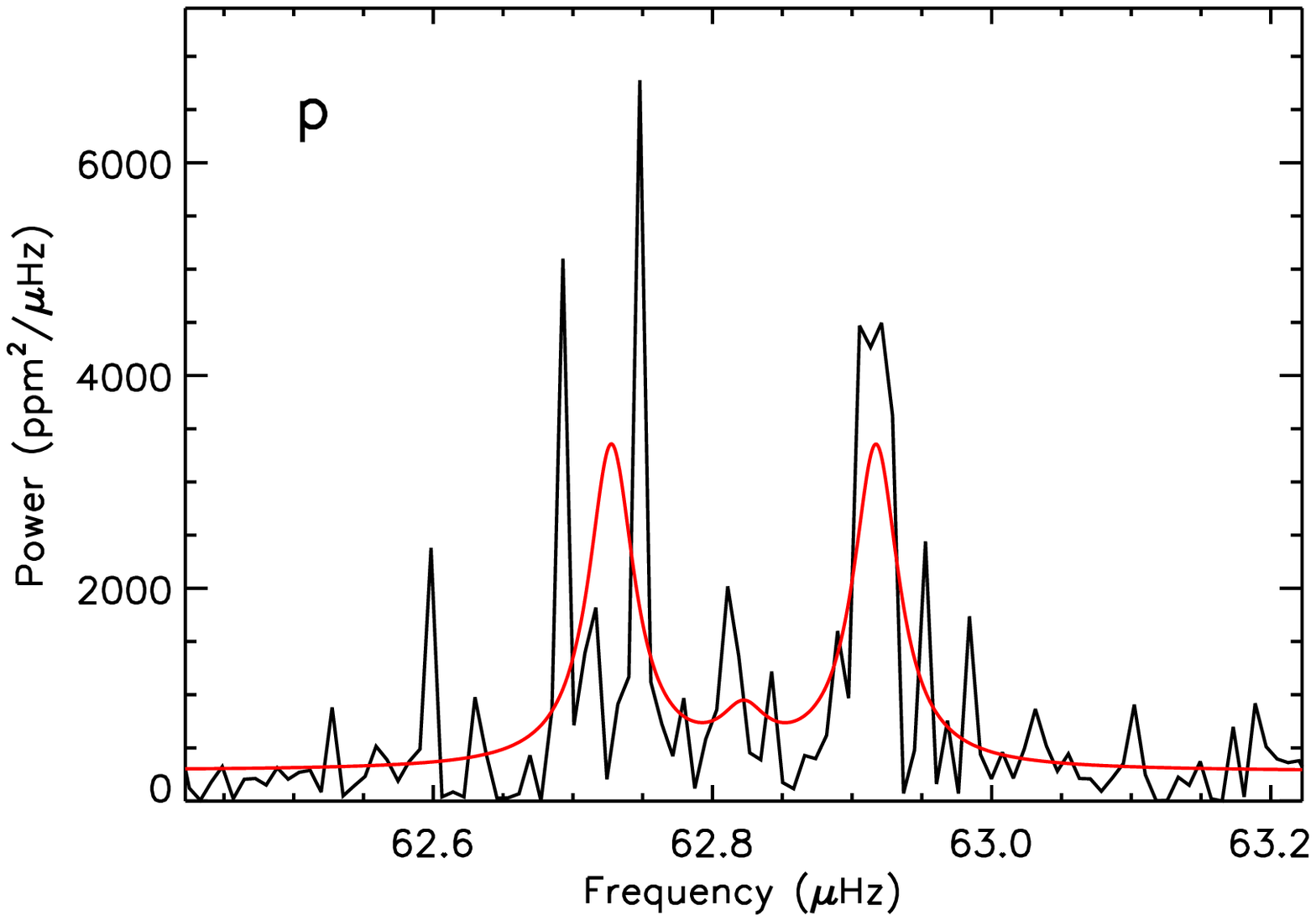}
\end{center}
\caption{Section of the oscillation spectrum of KIC3744681 in the neighborhood of 4 p-dominated modes. The red curves represent the best fits to the data.
\label{fig_pmodes}}
\end{figure*}

\begin{itemize*}
\item \underline{$H_0$ hypothesis}: the observed multiplet has a splitting much smaller than the mode width ($\delta\nu_{\rm s}\ll\Gamma$). In this case, the mean profile of the whole multiplet corresponds to a single Lorentzian profile and is described by $n_0=3$ parameters (frequency $\nu_0$ of the central component, linewidth $\Gamma$, and height $H$).
\item \underline{$H_1$ hypothesis}: the rotational splitting of the mode is at least of the order of magnitude of the mode width ($\delta\nu_{\rm s}\gtrsim\Gamma$). The mean profile then corresponds to a well separated multiplet and is described by $n_1=5$ parameters (same ones as for the $H_0$ hypothesis plus the inclination angle $i$ and the splitting $\delta\nu_{\rm s}$).
\end{itemize*}

Let us denote as $\ell_0$ and $\ell_1$ the likelihoods of the best fits obtained under the $H_0$ to the $H_1$ hypotheses, respectively. We note that fitting the power spectrum in the neighborhood of the considered multiplet under the $H_1$ hypothesis necessarily gives an agreement with the data at least as good as fitting under the $H_0$ hypothesis. Indeed, the best fit under the $H_0$ hypothesis can also be obtained with the $H_1$ hypothesis and a pole-on inclination ($i=0^\circ$). As a consequence, the likelihood $\ell_0$ is always lower than the likelihood $\ell_1$. To determine whether or not the $H_0$ hypothesis can be rejected, one needs to assess the significance of the likelihood improvement when switching from the $H_0$ to the $H_1$ hypothesis. \cite{wilks38} has shown that the quantity $\Delta\Lambda\equiv-2\left[\ln(\ell1) - \ln(\ell_0)\right]$ follows the distribution of a $\chi^2$ with $(n_1-n_0)$ degrees of freedom (see also \citealt{appourchaux98}). We have verified in our particular case that $\Delta\Lambda$ indeed follows the distribution of a $\chi^2$ with $n_1-n_0=2$ degrees of freedom by performing Monte Carlo simulations (see Appendix \ref{app_wilks}). 

Supposing that we have computed the value $\Delta\Lambda$ for a given mode, the false-alarm probability is estimated by computing the probability that a mode in the $H_0$ hypothesis (i.e. not rotationally split) can produce such a high value of $\Delta\Lambda$. This probability is given by
\begin{linenomath*}
\begin{equation}
p=P(\chi^2(2 \hbox{ degrees of freedom}) \geqslant\Delta\Lambda)
\label{eq_pval}
\end{equation}
\end{linenomath*}
and is commonly referred to as the $p$-value. Low $p$-values indicate that a mode described as a sing Lorentzian profile is unlikely to produce the observations.

For each observed $l=1$ mode, we computed the $p$-value using Eq. \ref{eq_pval}, and we rejected all the modes for which a $p$-value higher than 0.05 was found. We applied this process to all the targets that were pre-selected in Sect. \ref{sect_select}, and kept only the stars for which at least five significant splittings were obtained. This left us with seven targets, which are listed in Table \ref{tab_targets}, with a number of significant splittings ranging from 6 to 14. The extracted frequencies and rotational splittings of the selected multiplets are given along with their $p$-values in Tables \ref{tab_lkhd} and \ref{tab_lkhd2}  for all the stars of this sample. To each multiplet corresponds an estimate of the inclination angle $i$. We note that for every star the angles found for each mode agree quite well with one another and consistently point toward inclination angles close to 90$^\circ$ (see Tables \ref{tab_lkhd} and \ref{tab_lkhd2}). This is understandable considering that the stars were selected so that the modes are clearly split, which greatly favors stars with high inclination angles, as pointed out in Sect. \ref{sect_select}. We note that in two cases (mode at $\nu\sim 89.1\,\mu$Hz for KIC7581399 and mode at $\nu\sim 92.6\,\mu$Hz for KIC5184199), the best-fit solutions point toward a solution with an inclination angle around $i=55^\circ$ and a splitting around twice the value obtained for the other modes. This is likely caused by the influence of a higher-degree mixed mode nearby (see discussion about these modes in Sect.~\ref{sect_mcmc}), and these two modes were discarded in the following.


\subsection{Bayesian approach \label{sect_mcmc}}

The limits of the $H_0$ test used in the frequentist approach are well known. To summarize in our case, a low $p$-value ensures that noise (in our case a multiplet where the splitting is much smaller than the mode width) has little chances to produce the observed signal. However, it says nothing about the probability to produce the observations under the $H_1$ hypothesis (i.e. when the multiplet is detectably split by rotation).

This can be remedied by applying a Bayesian approach, which produces an estimate of the consistency between a given model $M_j$ and a dataset $D$. In the present case, the dataset $D$ corresponds to the observed power spectrum in the neighborhood of a given detected $l=1$ mode, and we aim at testing the two models of power spectrum $M_0$ and $M_1$ that correspond to the hypotheses $H_0$ and $H_1$ defined in Sect \ref{sect_MLE}. In other words, model $M_0$ is built assuming of a single Lorentzian profile, whereas model $M_1$ corresponds to a triplet of Lorentzian profiles. Each model $M_j$ relies on a set of free parameters $\boldsymbol{\theta}$, which contains for both models the background level, the central frequency $\nu_0$, the mode linewidth $\Gamma$, and the mode amplitude $A$, and additionally for model $M_1$ the inclination angle $i$ and the rotational splitting $\delta\nu_{\rm s}$. 

Within a Bayesian approach, we must define a prior probability law $\pi(\boldsymbol{\theta}|M_j,I)$, based on a certain number of assumptions on the signal that are here summarized as $I$. The chosen priors are explicited below. We used the Bayes theorem to define the posterior probability density function of a given set of parameters $\boldsymbol{\theta}$ assuming model $M_j$ and knowing the dataset $D$ through the relation
\begin{equation}
 \pi(\boldsymbol{\theta}|D,M_j,I)=\frac{\pi(\boldsymbol{\theta}|M_j,I)\pi(D|\boldsymbol{\theta},M_j,I)}{P(D|M_j,I)}
\end{equation}
where $\pi(D|\boldsymbol{\theta},M_j,I)$ corresponds to the likelihood already introduced in Sect. \ref{sect_MLE}, $\pi(\boldsymbol{\theta}|M_j,I)$ corresponds to the afore-mentioned priors, and $P(D|M,I)$ is the \textit{evidence} of model $M_j$ knowing the data $D$ and is defined as
\begin{equation}
 P(D|M_j,I)=\int \pi(\boldsymbol{\theta}|M_j,I)\pi(D|\boldsymbol{\theta},M_j,I) d\boldsymbol{\theta}.
\label{eq_evidence}
\end{equation}

\begin{table*}
  \begin{center}
  \caption{Extracted frequencies $\nu_0$ and rotational splittings $\delta\nu_{\rm s}$ for the $l=1$ modes of KIC8962923, KIC5184199, KIC4659821, and KIC3744681. The $3^{\rm rd}$ column gives the estimate of the inclination angle $i$ obtained from each mode. Only modes whose splitting was found significant with a $p$-value (given in $4^{\rm th}$ column) below 0.05 are given. The $5^{\rm th}$ column gives the posterior probability obtained from a Bayesian analysis that the rotational splitting of the mode is detected (see Sect. \ref{sect_mcmc}). The modes whose parameters are shown in italics were discarded (see text). \label{tab_lkhd}}

\textsc{KIC8962923} \\
\vspace{0.2cm}
\begin{tabular}{c c c c | c}
\hline \hline
\T \B $\nu_0$ ($\mu$Hz) & $\delta\nu_{\rm s}$ ($\mu$Hz) & $i$ ($^\circ$) & $p$-value & Posterior probability (\%) \\
\hline
\T $  57.275\pm0.004$ & $0.070\pm0.004$ & $70.68\pm  8.97$ & $0.015$ & $ 49.3$ \\
$  70.532\pm0.004$ & $0.069\pm0.004$ & $81.46\pm  6.65$ & $0.000$ & $ 99.1$ \\
$  71.883\pm0.006$ & $0.058\pm0.006$ & $90.00\pm 14.54$ & $0.004$ & $ 98.3$ \\
$  72.932\pm0.010$ & $0.080\pm0.010$ & $90.00\pm 10.66$ & $0.027$ & $ 91.2$ \\
$  73.958\pm0.004$ & $0.065\pm0.005$ & $68.92\pm  6.47$ & $0.001$ & $ 99.5$ \\
$  77.046\pm0.004$ & $0.077\pm0.004$ & $72.66\pm  6.30$ & $0.005$ & $ 97.5$ \\
$  78.672\pm0.006$ & $0.083\pm0.006$ & $65.72\pm  7.83$ & $0.020$ & $ 94.9$ \\
$  81.110\pm0.008$ & $0.056\pm0.008$ & $90.00\pm 15.31$ & $0.042$ & $ 67.9$ \\
\B $  84.792\pm0.007$ & $0.079\pm0.007$ & $80.62\pm  7.29$ & $0.001$ & $ 99.8$ \\
\hline 
\end{tabular}
\vspace{0.5cm}

\textsc{KIC5184199} \\
\vspace{0.2cm}
\begin{tabular}{c c c c | c}
\hline \hline
\T \B $\nu_0$ ($\mu$Hz) & $\delta\nu_{\rm s}$ ($\mu$Hz) & $i$ ($^\circ$) & $p$-value & Posterior probability (\%) \\
\hline
\T $  69.466\pm0.009$ & $0.105\pm0.009$ & $90.00\pm 9.60$ & $0.019$ & $ 92.0$ \\
$  76.319\pm0.008$ & $0.086\pm0.009$ & $90.00\pm 6.70$ & $0.003$ & $ 95.0$ \\
$  82.640\pm0.011$ & $0.098\pm0.011$ & $90.00\pm12.98$ & $0.039$ & $ 55.8$ \\
$  84.883\pm0.009$ & $0.080\pm0.009$ & $89.99\pm67.88$ & $0.008$ & $ 99.1$ \\
$  91.505\pm0.007$ & $0.093\pm0.007$ & $90.00\pm10.56$ & $0.000$ & $100.0$ \\
$  \mathit{92.614\pm0.009}$ & $\mathit{0.130\pm0.012}$ & $\mathit{54.09\pm 6.90}$ & $\mathit{0.010}$ & $\mathit{86.2}$ \\
$  94.115\pm0.010$ & $0.102\pm0.009$ & $90.00\pm 9.93$ & $0.000$ & $ 98.2$ \\
$  98.241\pm0.004$ & $0.093\pm0.004$ & $79.98\pm 4.10$ & $0.000$ & $100.0$ \\
$ 101.195\pm0.009$ & $0.076\pm0.010$ & $90.00\pm17.27$ & $0.003$ & $ 96.7$ \\
\B $ 105.753\pm0.014$ & $0.106\pm0.013$ & $90.00\pm17.29$ & $0.043$ & $ 99.6$ \\
\hline 
\end{tabular}
\vspace{0.5cm}

\textsc{KIC4659821} \\
\vspace{0.2cm}
\begin{tabular}{c c c c | c}
\hline \hline
\T \B $\nu_0$ ($\mu$Hz) & $\delta\nu_{\rm s}$ ($\mu$Hz) & $i$ ($^\circ$) & $p$-value & Posterior probability (\%) \\
\hline
\T $  80.659\pm0.007$ & $0.088\pm0.008$ & $74.27\pm 7.66$ & $0.002$ & $ 93.2$ \\
$  94.645\pm0.004$ & $0.082\pm0.005$ & $78.71\pm 5.16$ & $0.000$ & $ 99.8$ \\
$  96.193\pm0.009$ & $0.069\pm0.009$ & $90.00\pm16.60$ & $0.019$ & $ 91.7$ \\
$ 110.366\pm0.008$ & $0.085\pm0.008$ & $90.00\pm 9.88$ & $0.006$ & $ 99.5$ \\
$ 112.513\pm0.012$ & $0.082\pm0.017$ & $79.58\pm22.04$ & $0.011$ & $ 83.5$ \\
\B $ 113.765\pm0.010$ & $0.084\pm0.013$ & $71.38\pm10.90$ & $0.006$ & $ 94.8$ \\
\hline 
\end{tabular}
\vspace{0.5cm}

\textsc{KIC3744681} \\
\vspace{0.2cm}
\begin{tabular}{c c c c | c}
\hline \hline
\T \B $\nu_0$ ($\mu$Hz) & $\delta\nu_{\rm s}$ ($\mu$Hz) & $i$ ($^\circ$) & $p$-value & Posterior probability (\%) \\
\hline
\T $  51.747\pm0.002$ & $0.093\pm0.002$ & $70.30\pm  8.08$ & $0.000$ & $ 71.6$ \\
$  52.286\pm0.009$ & $0.108\pm0.010$ & $77.16\pm  6.49$ & $0.018$ & $ 75.5$ \\
$  57.558\pm0.010$ & $0.082\pm0.009$ & $90.00\pm  8.51$ & $0.015$ & $ 92.6$ \\
$  58.093\pm0.007$ & $0.082\pm0.007$ & $90.00\pm 17.37$ & $0.000$ & $ 96.4$ \\
$  61.856\pm0.004$ & $0.090\pm0.004$ & $73.38\pm  9.17$ & $0.000$ & $ 99.3$ \\
$  62.822\pm0.007$ & $0.095\pm0.007$ & $79.48\pm  5.76$ & $0.000$ & $ 99.7$ \\
$  63.542\pm0.009$ & $0.088\pm0.010$ & $74.02\pm  8.09$ & $0.000$ & $ 97.8$ \\
$  64.215\pm0.009$ & $0.109\pm0.009$ & $80.26\pm  8.43$ & $0.001$ & $ 93.3$ \\
$  67.845\pm0.009$ & $0.097\pm0.009$ & $90.00\pm142.99$ & $0.000$ & $ 98.8$ \\
$  68.803\pm0.010$ & $0.074\pm0.011$ & $80.16\pm 12.18$ & $0.017$ & $ 70.8$ \\
$  69.560\pm0.011$ & $0.121\pm0.011$ & $69.23\pm  7.10$ & $0.020$ & $ 72.6$ \\
$  73.660\pm0.008$ & $0.088\pm0.007$ & $90.00\pm 11.93$ & $0.000$ & $ 92.3$ \\
$  74.545\pm0.009$ & $0.100\pm0.010$ & $67.17\pm  7.29$ & $0.004$ & $ 58.8$ \\
\B $  75.537\pm0.013$ & $0.098\pm0.011$ & $79.17\pm 12.22$ & $0.036$ & $ 69.1$ \\
\hline 
\end{tabular}

\end{center}
\end{table*}

\begin{table*}
  \begin{center}
  \caption{Same as Table \ref{tab_lkhd} for targets KIC7467630, KIC9346602, and KIC7581399. \label{tab_lkhd2}}

\textsc{KIC7467630} \\
\vspace{0.2cm}
\begin{tabular}{c c c c | c}
\hline \hline
\T \B $\nu_0$ ($\mu$Hz) & $\delta\nu_{\rm s}$ ($\mu$Hz) & $i$ ($^\circ$) & $p$-value & Posterior probability (\%) \\
\hline
\T $  61.557\pm0.002$ & $0.059\pm0.001$ & $58.65\pm47.98$ & $0.032$ & $ 60.7$ \\
$  67.666\pm0.004$ & $0.064\pm0.003$ & $89.99\pm16.34$ & $0.000$ & $ 59.0$ \\
$  68.903\pm0.008$ & $0.064\pm0.007$ & $90.00\pm13.72$ & $0.011$ & $ 97.1$ \\
$  69.891\pm0.010$ & $0.076\pm0.010$ & $89.99\pm23.16$ & $0.005$ & $ 94.6$ \\
$  70.769\pm0.008$ & $0.071\pm0.008$ & $90.00\pm20.24$ & $0.003$ & $ 98.7$ \\
\B $  82.877\pm0.010$ & $0.101\pm0.012$ & $69.44\pm 7.59$ & $0.011$ & $ 70.3$ \\
\hline 
\end{tabular}
\vspace{0.5cm}

\textsc{KIC9346602} \\
\vspace{0.2cm}
\begin{tabular}{c c c c | c}
\hline \hline
\T \B $\nu_0$ ($\mu$Hz) & $\delta\nu_{\rm s}$ ($\mu$Hz) & $i$ ($^\circ$) & $p$-value & Posterior probability (\%) \\
\hline
\T $  43.638\pm0.005$ & $0.073\pm0.005$ & $82.85\pm 7.72$ & $0.000$ & $ 96.4$ \\
$  44.324\pm0.005$ & $0.067\pm0.005$ & $90.00\pm10.53$ & $0.000$ & $ 98.6$ \\
$  48.203\pm0.006$ & $0.083\pm0.006$ & $90.00\pm14.14$ & $0.001$ & $ 91.8$ \\
$  48.688\pm0.007$ & $0.070\pm0.007$ & $78.18\pm 8.28$ & $0.007$ & $ 70.2$ \\
$  53.694\pm0.004$ & $0.080\pm0.004$ & $90.00\pm 4.77$ & $0.000$ & $100.0$ \\
$  54.119\pm0.005$ & $0.074\pm0.005$ & $82.59\pm 6.82$ & $0.000$ & $ 99.8$ \\
$  54.678\pm0.004$ & $0.077\pm0.005$ & $78.11\pm 5.24$ & $0.000$ & $ 99.4$ \\
$  57.587\pm0.007$ & $0.098\pm0.008$ & $76.24\pm 6.34$ & $0.023$ & $ 78.7$ \\
$  58.334\pm0.007$ & $0.073\pm0.007$ & $90.00\pm 5.94$ & $0.006$ & $ 91.6$ \\
$  60.037\pm0.003$ & $0.071\pm0.004$ & $77.90\pm 5.35$ & $0.019$ & $ 58.1$ \\
$  62.775\pm0.003$ & $0.080\pm0.003$ & $72.70\pm 7.46$ & $0.000$ & $ 99.4$ \\
$  64.942\pm0.008$ & $0.074\pm0.008$ & $90.00\pm17.27$ & $0.038$ & $ 65.6$ \\
$  67.956\pm0.005$ & $0.082\pm0.005$ & $81.02\pm 7.68$ & $0.000$ & $ 90.1$ \\
\B $  73.922\pm0.002$ & $0.079\pm0.003$ & $78.09\pm 6.60$ & $0.000$ & $ 97.5$ \\
\hline 
\end{tabular}
\vspace{0.5cm}

\textsc{KIC7581399} \\
\vspace{0.2cm}
\begin{tabular}{c c c c | c}
\hline \hline
\T \B $\nu_0$ ($\mu$Hz) & $\delta\nu_{\rm s}$ ($\mu$Hz) & $i$ ($^\circ$) & $p$-value & Posterior probability (\%) \\
\hline
\T $  63.546\pm0.006$ & $0.084\pm0.006$ & $74.69\pm 7.28$ & $0.005$ & $95.7$ \\
 $  67.852\pm0.006$ & $0.086\pm0.004$ & $90.00\pm10.37$ & $0.001$ & $50.5$ \\
 $  68.811\pm0.005$ & $0.079\pm0.005$ & $80.13\pm 8.35$ & $0.031$ & $49.7$ \\
 $  76.375\pm0.010$ & $0.083\pm0.011$ & $78.07\pm12.43$ & $0.003$ & $88.8$ \\
 $  81.297\pm0.005$ & $0.082\pm0.005$ & $90.00\pm16.03$ & $0.000$ & $92.3$ \\
 $  82.502\pm0.009$ & $0.072\pm0.012$ & $78.81\pm15.49$ & $0.023$ & $89.4$ \\
 $  83.314\pm0.010$ & $0.095\pm0.009$ & $89.98\pm19.56$ & $0.001$ & $98.4$ \\
\B $  \mathit{89.056\pm0.010}$ & $\mathit{0.155\pm0.011}$ & $\mathit{60.01\pm 6.93}$ & $\mathit{0.001}$ & $\mathit{94.7}$ \\
\hline 
\end{tabular}

\end{center}
\end{table*}

We assumed the following prior probability laws for the free parameters $\boldsymbol{\theta}$:
\begin{itemize}
 \item amplitude $A$: uniform prior for the variable $\ln\pi A^2$ over $[\ln(P_0/1000),\ln(1.2 P_0)]$ where $P_0$ is the total power in the fitting window. This non-informative prior is very loose.
 \item width $\Gamma$: uniform prior for $\ln \Gamma$ over the range $[\ln r,\ln(0.4\ \mathrm{\mu Hz})]$ with Gaussian decays around these boundaries. $r$ is the frequency resolution of the spectrum (about $0.008\,\mu$Hz) and $0.4\ \mathrm{\mu Hz}$ corresponds to an upper limit of radial mode widths observed in these stars. With this prior, we assumed that the widths of $l=1$ mixed modes cannot excess those of $l=0$ modes, which are purely acoustic.
 \item central frequency $\nu_0$: uniform prior for $\nu_0$ spanning an interval of 0.6~$\mu$Hz around the middle of the fitting window.
 \item rotation splitting $\delta\nu_s$: uniform prior for $\delta\nu_s$ over [0,0.3~$\mathrm{\mu Hz}$].
 \item inclination angle $i$: uniform prior for $\cos i$ over [0,1]. This prior derives from the assumption of isotropy of the inclinations of stars.
 \item background level: we used as a prior the result of a global fit of the background as described in Sect. \ref{sect_MLE}. By doing so, we used information coming from outside the fitting window.
\end{itemize}

To compare two different models $M_i$ and $M_j$ we used the evidence of these models as introduced in Eq. \ref{eq_evidence} to define the so-called \textit{odd ratio}
\begin{equation}
 O_{ij}=\frac{P(M_i|D,I)}{P(M_j|D,I)}=\frac{P(M_i|I)}{P(M_j|I)}\frac{P(D|M_i,I)}{P(D|M_j,I)}.
 \label{eq_oddratio}
\end{equation}
By assuming that the two competing models $M_i$ and $M_j$ are equiprobable, i.e. $P(M_i|I)=P(M_j|I)=1/2$, the odd ratio is equivalent to the Bayes factor
\begin{equation}
 O_{ij}=B_{ij}=\frac{P(D|M_i,I)}{P(D|M_j,I)}.
\end{equation}
The relative probability to favor model $M_j$ over $M_i$ is then given as
\begin{equation}
P(M_j|D)=(1+O_{ij})^{-1}
\end{equation}
which derives from Eq. \ref{eq_oddratio} if we assume that no other model than models $M_i$ and $M_j$ is possible.

To calculate the evidences of models $M_0$ and $M_1$, we developed an automated parallel tempering Markov-Chain Monte Carlo similar to the one developed by \citet{benomar09a}. We then computed the odd ratio $O_{01}$ for all the modes that were found to have a $p$-value below 0.05 in Sect. \ref{sect_MLE}, and derived the probability $P(M_1|D)$ to favor model $M_1$. The results are listed in the last column of Tables \ref{tab_lkhd} and \ref{tab_lkhd2}. All the modes for which the obtained probability was higher than 85\% were considered as reliable and used to probe the rotation profiles of the selected targets.

We stress that the approach followed here is conservative. In particular, we have assumed that either model $M_0$ or model $M_1$ is correct, i.e. that the contribution from only one $l=1$ mode is present in the fitting window. If other non-negligible components (such as higher degree modes) are also present then model $M_1$ can appear nearly as bad as model $M_0$ in describing the spectrum, even though a clear $l=1$ multiplet is present. This occurs especially for g-dominated modes, which lie in a region of the spectrum where mixed $l=2$ and $l=3$ modes can be present. For these modes, the relatively low posterior probability does not mean that the splitting cannot be measured, but that the model would need to be improved. In the present case, we did not seek to improve it because enough g-dominated modes were found significant with the current approach to estimate the amount of differential rotation (see Sect. \ref{sect_interpretation}).

By imposing that the measured splittings should produce both a low $p$-value (frequentist approach) and a high posterior probability $P(M_1|D)$ (Bayesian approach), we ensured that only very robust splitting detections are validated, at the expense of potentially discarding good data. However, a robust selection process was needed to clearly establish that the splittings of both p and g-dominated modes can be reliably extracted, which is a necessary condition to estimate the amount of differential rotation. With this double selection process, we retained between 3 and 10 significant splittings for the stars of the sample. In the following only these splittings are used to infer the internal rotation profile, but the splittings that passed only the frequentist test are also shown on the plots.

\section{Measuring the amount of differential rotation \label{sect_interpretation}}

Two different approaches have been proposed to obtain information on the internal rotation profile of red giants using the mode splittings that were selected in Sect. \ref{sect_analysis}. 
\begin{enumerate}
\item The first one requires to obtain a stellar model that closely matches the observed mode frequencies, as well as the classical constraints on the star (atmospheric parameters). If such a model can be found, one gains access to the eigenfunctions of the modes, which makes it possible to determine where the modes are trapped and to compute the rotational kernels of the modes. Information on the rotation profile can then be obtained either by performing inversions of Eq. \ref{eq_splitting} (D12, D14), or by applying forward techniques (\citealt{beck14}). In spite of these advantages, this method is time-consuming because it requires to compute a model that matches the observed mode frequencies. More importantly, it is model-dependent. This can be a problem here since the high values of $\Delta\Pi_1$ that are observed for a good proportion of \kepler\ clump stars fail to be reproduced by several evolution codes (e.g. \citealt{montalban13}), although some codes might not have this problem (\citealt{lagarde12}). It was suggested that this discrepancy could be solved by adding overshooting at the boundary of the convective core induced by He-burning (\citealt{montalban13}), but this is still a matter of debate.
\vspace{0.2cm}
\item A simplified approach was proposed by G13. By using asymptotic developments from \cite{unno89}, they showed that estimates of the mean rotation rate in the g-mode and p-mode cavities can be obtained using only the mode frequencies, i.e. without having to compute a stellar model of the observed target. This constitutes an appealing alternative to the first method in the cases where the observed $\Delta\Pi_1$ cannot be reproduced by stellar models. The downside is that it relies on several layers of approximations and needs to be tested in the case of clump stars.
\end{enumerate}

We applied both approaches to one test-target of our sample. We chose \cible\ because it has a value of $\Delta\Pi_1$ that is low enough to be satisfactorily reproduced by current stellar models, so that the first approach can be applied. We note that the internal structure of the best-fit models obtained for \cible\ (see \ref{sect_inversions}) is probably not entirely correct, because we know that current models underestimate the values of $\Delta\Pi_1$ of clump stars. However, it has been shown that the rotation profiles obtained from inversions do not critically depend on the input physics of the reference model, provided the frequencies of the observed modes are well reproduced (D12, D14). Our objectives were twofold. First, we wanted to verify that the rotational splittings that were extracted in Sect. \ref{sect_analysis} are sufficient to provide a constraint on the ratio between the core and the envelope rotation rates. Secondly, we aimed at testing the simplified method of G13 on red clump stars, in order to determine whether it can be applied to all targets.

\subsection{Model-dependent method: the test case of \cible \label{sect_inversions}}

\subsubsection{Modeling \cible\ \label{sect_model}}

\begin{figure}
\begin{center}
\includegraphics[width=9cm]{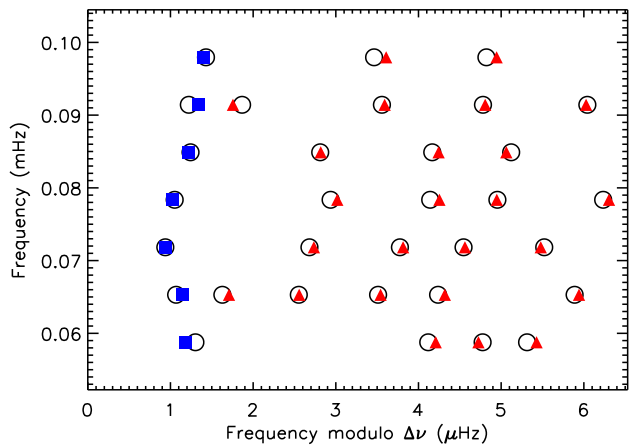}
\end{center}
\caption{Echelle diagram of the best-fit model for \cible. The open circles correspond to the observed frequencies and the colored filled symbols represent those of the model (blue squares: $l=0$ modes, red triangles: $l=1$ modes).
\label{fig_echelle_bestmod}}
\end{figure}

In order to perform inversions of Eq. \ref{eq_splitting}, one needs to have access to the rotational kernels of the modes, and therefore to a stellar model of the star. For this purpose, we used the evolutionary code MESA (\citealt{mesa}). We used the OPAL 2005 equation of state (\citealt{rogers02}) and opacity tables. The nuclear reaction rates were computed using the NACRE compilation (\citealt{angulo99}). We assumed the classical solar mixture of heavy elements of \cite{grevesse93}. Convection was described using the classical mixing length theory (\citealt{bohm58}) with a mixing length parameter calibrated on the Sun ($\alpha_{\rm MLT} = 2.08$). The effects of microscopic diffusion and overshooting from convective regions were neglected in this study.

To model \cible, we followed the procedure proposed by \cite{deheuvels11} to model subgiants with mixed modes, and later adapted to red giants by D12. We refer to these papers for more details about the method, which is only briefly recalled here. The authors showed that the combined knowledge of the mean large separation $\langle\Delta\nu\rangle$ and mean period spacing of $l=1$ modes $\langle\Delta\Pi_1\rangle$ yields tight constraints on the stellar mass and age when other physical inputs are fixed. To model a red giant, they thus advocated to vary all physical parameters except for the mass and age, which are determined each time through an iterative process so that the observed $\langle\Delta\nu\rangle$ and $\langle\Delta\Pi_1\rangle$ are correctly reproduced. In practice, the constraint given by $\langle\Delta\nu\rangle$ is advantageously replaced by the observed frequencies of the lowest-order radial modes, because the latter are less sensitive to near-surface effects. For the models, the mean period spacing $\langle\Delta\Pi_1\rangle$ was estimated from the approximate expression
\begin{linenomath*}
\begin{equation}
\langle\Delta\Pi_1\rangle \approx \pi^2\sqrt{2}\left(\int_{r_{\rm a}}^{r_{\rm b}} \frac{N_{\rm BV}}{r}\,\hbox{d}r \right)^{-1}
\label{eq_deltapi}
\end{equation}
\end{linenomath*}
We note that $\langle\Delta\Pi_1\rangle$ could in principle be estimated from the mode frequencies themselves by following the procedure described in Sect. \ref{sect_deltapi}. This would however require to compute the frequencies of $l=1$ modes for a large number of models, which is time-consuming. It was shown that there is a good agreement with the estimate of $\langle\Delta\Pi_1\rangle$ obtained from Eq. \ref{eq_deltapi} and that obtained from the mode frequencies (\citealt{mosser12a}, D14).

We computed a grid of models, varying both the initial helium content $Y_{\rm i}$ and the initial metallicity $(Z/X)_{\rm i}$. For each point of this grid, the procedure described above was applied to estimate the mass and age. The mode frequencies of each model were then computed using the oscillation code \losc\ (\citealt{losc}). Before comparing them to the observed frequencies, we corrected them from the well-known near-surface effects using the recipe proposed by \cite{kjeldsen08}, which consists of adding to the mode frequencies a power law whose exponent is calibrated in the Sun. As was pointed out by \cite{kjeldsen08}, mixed modes have a larger inertia in the core than pure acoustic modes. Therefore, for these modes the correction terms were weighted by the ratio between the inertia of the closest radial mode to the inertia of the mode itself (\citealt{aerts10}).

We found a model that reproduces the surface parameters of the stars and closely matches the observed mode frequencies and the star's effective temperature. An \'echelle diagram of its mode frequencies corrected from near-surface effects as described above is shown in Fig. \ref{fig_echelle_bestmod}. This model, whose main properties are listed in Table \ref{tab_model}  is used as a reference model in the following.

\begin{table*}
  \begin{center}
  \caption{Parameters of the best-fit model of \cible \label{tab_model}}
\begin{tabular}{c c c c c c}
\hline \hline
\T \B Mass ($M_\odot$) & Age (Myr) & $Y_{\rm i}$ & $(Z/X)_{\rm i}$ & Radius ($R_\odot$) & $T_{\rm eff}$ (K) \\
\hline 
\T \B 2.65 & 5686 & 0.29 & 0.035 & 10.38 & 5129 \\  
\hline
\end{tabular}
\end{center}
\end{table*}

\subsubsection{Mode trapping \label{sect_trapping}}

\begin{figure}
\begin{center}
\includegraphics[width=9cm]{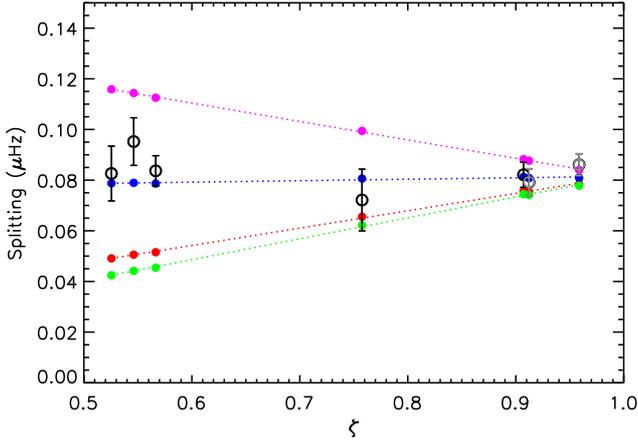}
\end{center}
\caption{Extracted rotational splittings for $l=1$ modes of \cible\ (open black circles indicate modes that passed both frequentist and Bayesian significance tests and open gray circles modes that passed the frequentist test only). The colored circles indicate theoretical splittings for linear rotation profiles that reproduce the rotation of \cible\ in the g-mode cavity as obtained from the OLA method, and with ratios $\Omega_{\rm core}/\Omega_{\rm surf}$ of 1 (purple), 2 (blue), 10 (red), or 100 (green) from top to bottom.
\label{fig_zeta_split}}
\end{figure}

The eigenfunctions of our reference model of \cible\ were used to estimate the trapping of the modes. For each mode, we computed the parameter $\zeta$ defined as the ratio between the kinetic energy of the mode in the g-mode cavity and the total kinetic energy
\begin{linenomath*}
\begin{equation}
\zeta \equiv \frac{I_{\rm g}}{I} = \frac{\int_{r\ind{a}}^{r\ind{b}} \rho r^2 \left[\xi_r^2+l(l+1)\xi_h^2\right]\,dr}{\int_0^{R_\star} \rho r^2 \left[\xi_r^2+l(l+1)\xi_h^2\right]\,dr},
\label{eq_zeta}
\end{equation}
\end{linenomath*}
where $\xi_r$ and $\xi_h$ are the radial and horizontal displacements. A value of $\zeta$ close to 1 indicates that the mode is mainly trapped in the g-mode cavity, and a value of $\zeta$ close to 0, that it is trapped in the p-mode cavity\footnote{The p-mode cavity depends on the degree $l$ of the mode. Since we here consider $l=1$ modes only, the p-mode cavity refers to the cavity of $l=1$ p modes throughout the paper.}. Fig. \ref{fig_zeta_split} represents the rotational splittings of \cible\ that passed the statistical test of significance described in Sect. \ref{sect_analysis} as a function of the mode trapping $\zeta$. We clearly observe that the splittings of p-dominated modes are comparable to the splittings of the g-dominated modes. This is a striking difference with the RGB stars, for which p-dominated modes have a splitting about half that of g-dominated modes (\citealt{mosser12b}). This already suggests that there might be less differential rotation in secondary clump stars than in RBG stars.

\subsubsection{Rotation in the g-mode cavity \label{sect_core}}

\begin{figure}
\begin{center}
\includegraphics[width=9cm]{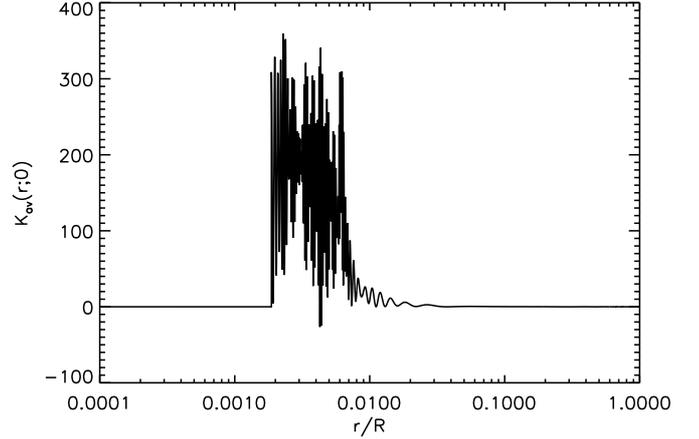}
\end{center}
\caption{Averaging kernel obtained for \cible\ with the OLA method in the g-mode cavity.
\label{fig_core_kernel}}
\end{figure}

It was shown for RGB stars that the mean rotation rate in the g-mode cavity can be precisely determined using the splittings of mixed modes (D12, \citealt{mosser12b}). For this purpose it is very convenient to use the OLA (optimally localized averages) method to invert Eq. \ref{eq_splitting}. This method, which was already successfully applied to the Sun (\citealt{schou98}, \citealt{chaplin99}) as well as to subgiants and RGB stars (D12, D14), consists of building combinations of the rotational kernels such that the resulting averaging kernels $\mathcal{K}(r;r_0)=\sum_k c_k(r_0)K_k(r)$ are as localized as possible around a target point $r_0$. When setting $r_0=r_{\rm g}$, where $r_{\rm g}$ is the mean radius of the g-mode cavity, we obtain an averaging kernel that is well localized in the g-mode cavity (see Fig. \ref{fig_core_kernel}). In particular, the contribution from the envelope is efficiently suppressed. This can be used to obtain a precise estimate of the mean rotation in the g-mode cavity $\langle\Omega_{\rm g}\rangle$. For \cible, we obtained $\langle\Omega_{\rm g}\rangle/(2\pi)=173\pm20$ nHz. 

We must stress that contrary to RGB stars, the core of clump stars is convective, due to He-burning. The convective core is very small because of the high temperature sensitivity of the 3$\alpha$ reaction (it represents less than 0.2\% of the stellar radius and less than 4\% of the stellar mass for \cible) but nonetheless, the most central layers are not probed by g modes, which are evanescent in convective regions. So we must keep in mind that the rotation in the very core cannot be probed by seismology.

\subsubsection{Core-envelope contrast}

One crucial point in this study, which aims at estimating the amount of radial differential rotation in clump stars, is to determine whether or not we can also obtain an estimate of the mean envelope rotation rate. 
For this purpose, we considered simple "two-zone" models that rotate as solid bodies with a rate $\Omega_{\rm g}$ in the g-mode cavity, and with a rate $\Omega_{\rm p}$ in the p-mode cavity. Naturally, we used for $\Omega_{\rm g}$ the estimate that was obtained from the observations in Sect. \ref{sect_core}. We chose values of $\Omega_{\rm p}$ such that the ratio $\Omega_{\rm g}/\Omega_{\rm p}$ is equal to 1, 2, 10, and 100 successively. These synthetic rotation profiles along with the rotational kernels of the best-fit model for \cible\ were plugged into Eq. \ref{eq_splitting} to obtain theoretical splittings, which are plotted as a function of the mode trapping $\zeta$ in Fig. \ref{fig_zeta_split}. This figure clearly shows that the splittings of p-dominated modes can be used to constrain the mean rotation in the p-mode cavities. For \cible, a SB rotation profile throughout the star ($\Omega_{\rm g}=\Omega_{\rm p}$) can be excluded, as well as rotation profiles with a large amount differential rotation ($\Omega_{\rm g}/\Omega_{\rm p}\geqslant 10$). The observations appear to be consistent with a core-envelope rotation contrast around 2 (blue curve).

This can be further quantified by searching for the parameters $(\Omega_{\rm g}, \Omega_{\rm p})$ that reproduce the observed splittings at best. For this purpose, we minimized the  reduced $\chi^2_{\rm red}$ defined as
\begin{linenomath*}
\begin{equation}
\chi^2\ind{red} = \frac{1}{M-2} \sum_{k=1}^M \left[\frac{\delta\nu_k-\Omega_{\rm p}A_k(r_0)-\Omega_{\rm g} B_k(r_0)}{\sigma_k}\right]^2,
\label{eq_chi2_2zone}
\end{equation}
\end{linenomath*}
where $\delta\nu_k$ are the $M$ observed splittings, $\sigma_k$ the corresponding error bars, $r_0$ is the limit between the p-mode and g-mode cavities, $A_k(r_0)\equiv\int_0^{r_0} K_k(r)\,\hbox{d}r$ and $B_k(r_0)\equiv\int_{r_0}^R K_k(r)\,\hbox{d}r$. We note that $r_0$ can be chosen anywhere in the evanescent zone between the p-mode and g-mode cavities with very little change in the results. For \cible, we obtained $\Omega_{\rm g}/(2\pi) = 156\pm12$ nHz and $\Omega_{\rm p}/(2\pi) = 95\pm15$ nHz. The mean rotation rate in the g-mode cavity is fully consistent with the result obtained with the OLA method. We thus obtained a core-envelope rotation contrast of $1.6\pm0.4$ for \cible. This shows that the amount of differential rotation in the secondary clump star \cible\ is much lower than for RGB stars.


\subsection{Testing the \cite{goupil13} approach on \cible \label{sect_G13}}

We then tested the simplified model-independent approach proposed by G13 on our test-target \cible\ in order to determine whether it can be applied to the targets of our sample. The method of G13 relies on two consecutive approximations.

\begin{figure}
\begin{center}
\includegraphics[width=9cm]{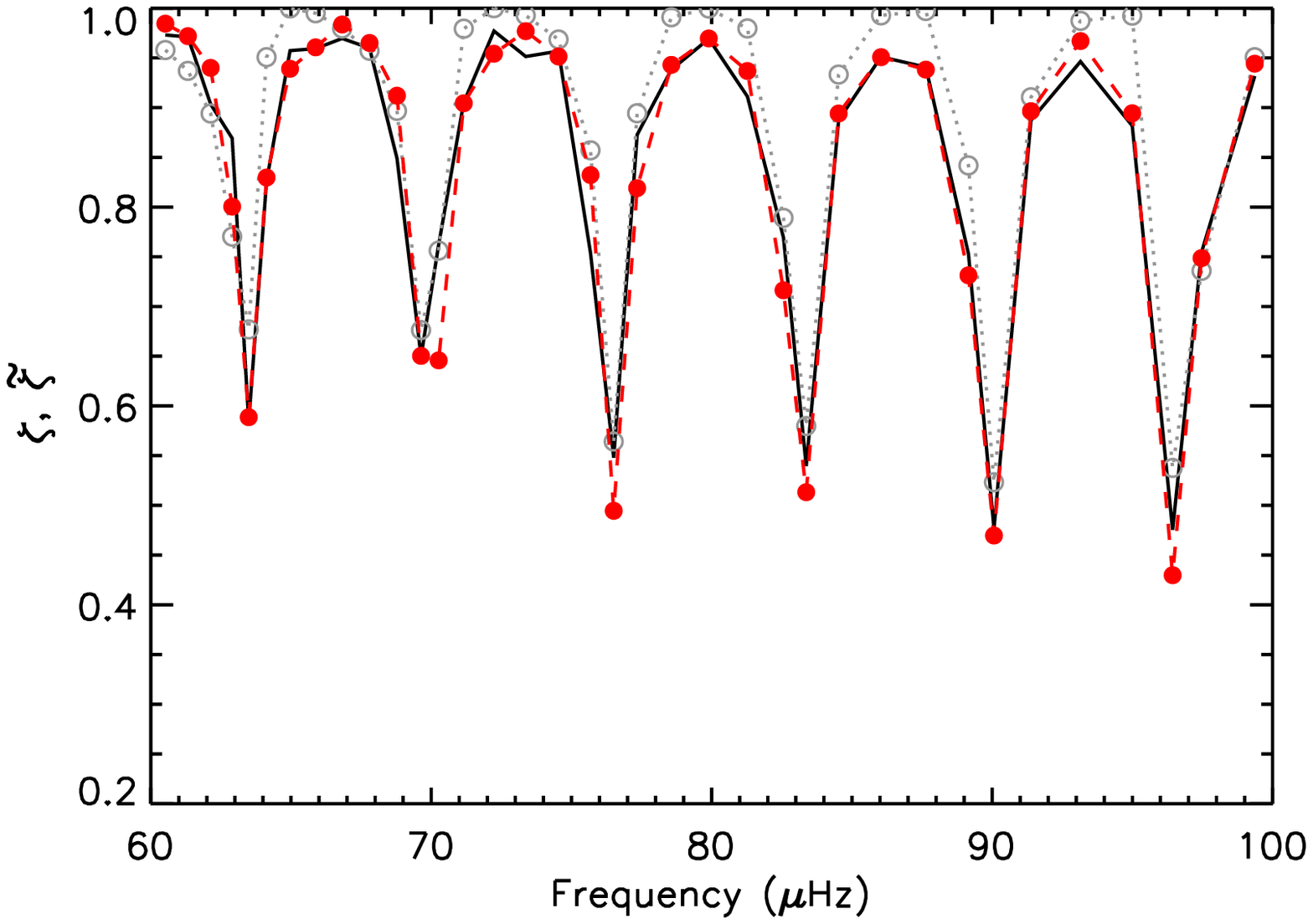}
\end{center}
\caption{Comparison between the mode trapping parameter $\zeta$ of the $l=1$ modes of the best-fit model for \cible\ computed from the mode eigenfunctions (Eq. \ref{eq_zeta}, black solid line) and the approximate expression $\tilde{\zeta}$ computed from the mode frequencies using the prescription of G13 modified in Appendix \ref{app_G13} (Eq. \ref{eq_zetatilde}, red filled circles and red dashed line). For reference, we also show the $\tilde{\zeta}$ computed with the original prescription of G13 (gray dotted line and open circles).
\label{fig_compare_zeta}}
\end{figure}

The first step consists in approximating the mode trapping parameter $\zeta$ by an expression $\tilde{\zeta}$ that depends solely on the mode frequencies, with the advantage that $\tilde{\zeta}$ can be estimated from the observations. To establish this relation, the authors used approximate expressions of the eigenfunctions derived from JWKB analysis. For all the modes whose splittings could be reliably extracted in \cible, we computed $\tilde{\zeta}$ following the prescription of G13 (Eq. A.27 and A.28 of their paper). As can be seen in Fig. \ref{fig_compare_zeta} (gray symbols), we obtained a quite good agreement with the values of $\zeta$ computed from the mode eigenfunctions in Sect. \ref{sect_trapping}, at the exception of nearly pure-g modes for which $\tilde{\zeta}$ slightly overestimates $\zeta$. As described in Appendix \ref{sect_G13}, we found that the agreement could be improved by releasing one of the approximations made by G13. We therefore propose a slightly modified expression for $\tilde{\zeta}$ given by Eq. \ref{eq_zetatilde} (see Appendix \ref{app_G13} for more detail). The $\tilde{\zeta}$ computed with this latter expression are in excellent agreement with the actual $\zeta$ parameters (red symbols in Fig. \ref{fig_compare_zeta}). This new expression for $\tilde{\zeta}$ was tested on several other models of clump stars and systematically gave an agreement as good as the one shown in Fig. \ref{fig_compare_zeta}. We thus confirm that an estimate of the mode trapping can be obtained from the frequency modes without having to compute a seismic model of the star, as was claimed by G13.


\begin{table}
  \begin{center}
  \caption{Mean rotation rates in the g- and p-mode cavities obtained for the selected targets with the method of G13 \label{tab_other_stars}}
\begin{tabular}{l c c c}
\hline \hline
\TT \BB KIC-Id & $ \displaystyle \frac{ \langle\Omega_{\rm g}\rangle}{2\pi}$ (nHz) & $ \displaystyle \frac{\langle\Omega_{\rm p}\rangle}{2\pi}$ (nHz) &  $ \displaystyle \frac{\langle\Omega_{\rm g}\rangle}{\langle\Omega_{\rm p}\rangle}$ \\
\hline 
\T 8962923 & $138\pm8$ & $79\pm10$ & $1.8\pm0.3$ \\
5184199 & $200\pm13$ & $63\pm20$ & $3.2\pm1.0$ \\
4659821 & $165\pm14$ & $79\pm15$ & $2.1\pm0.4$ \\
3744681 & $194\pm20$ & $63\pm36$ & $3.1\pm1.8$ \\
7467630 & $121\pm18$ & $96\pm28$ & $1.3\pm0.4$ \\
9346602 & $164\pm6$ & $53\pm15$ & $3.1\pm0.9$ \\
\B 7581399 & $167\pm11$ & $84\pm14$ & $2.0\pm0.4$ \\
\hline
\end{tabular}
\end{center}
\end{table}

\begin{figure*}
\begin{center}
\includegraphics[width=7.8cm]{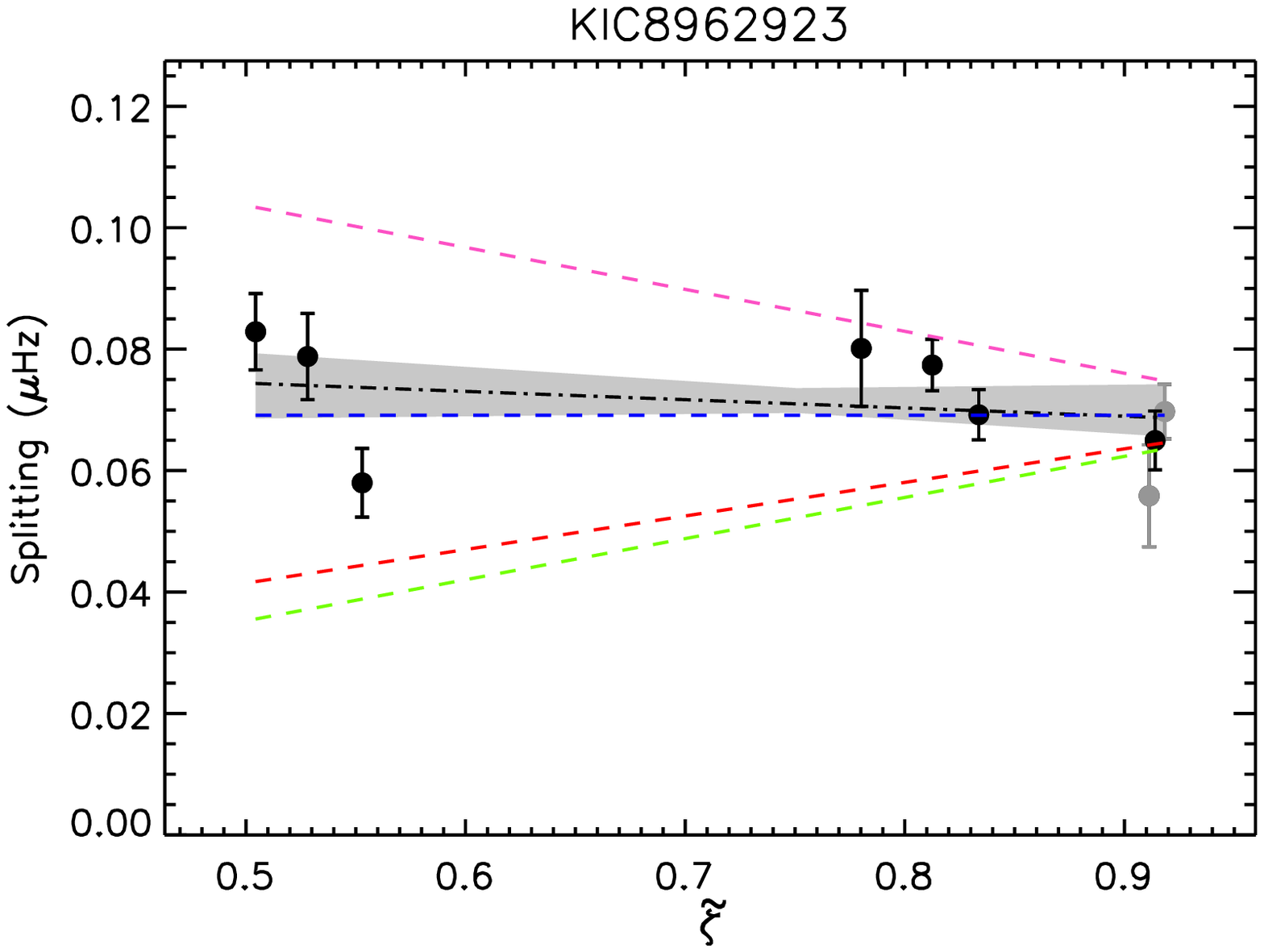}
\includegraphics[width=7.8cm]{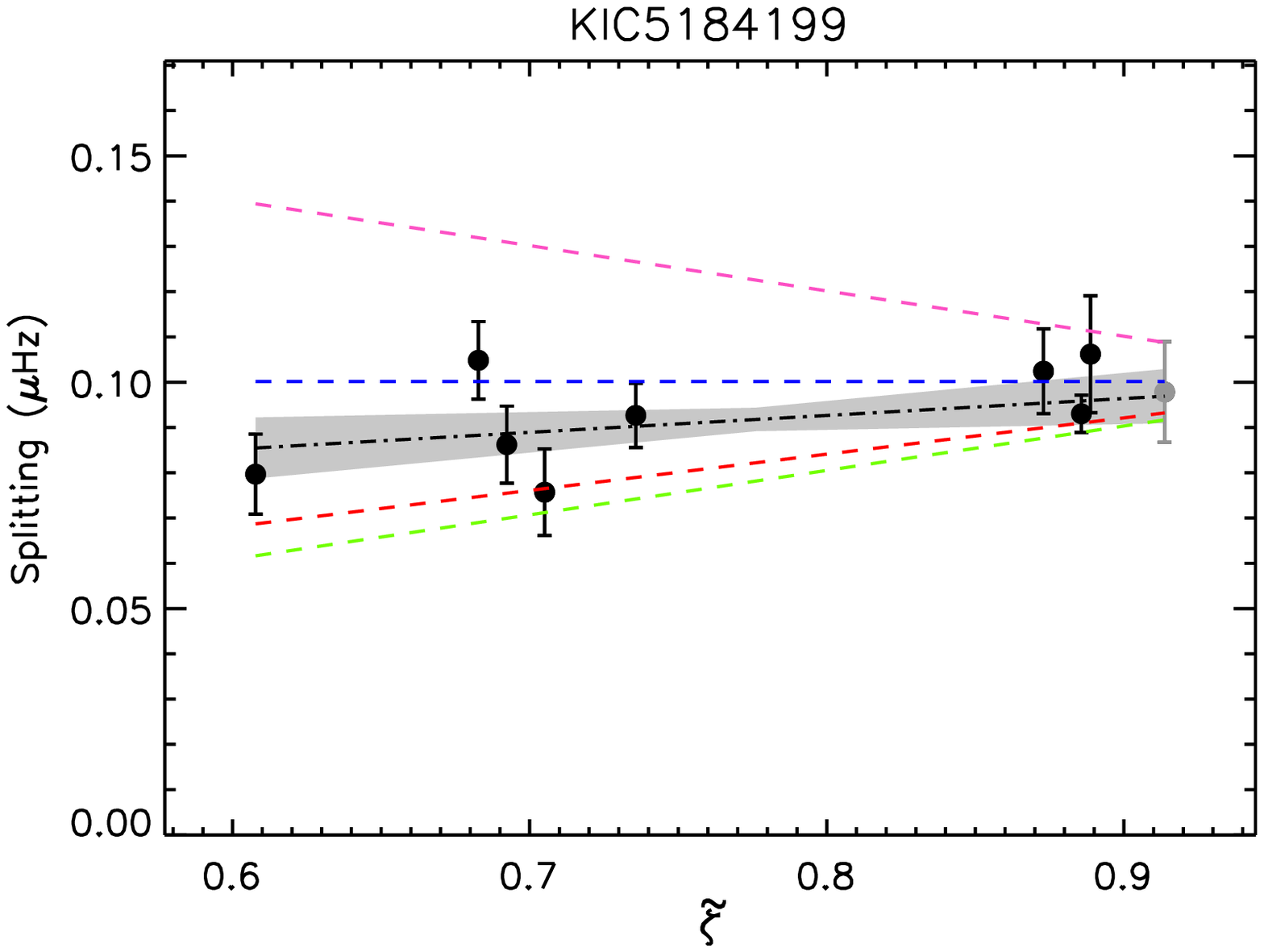}
\includegraphics[width=7.8cm]{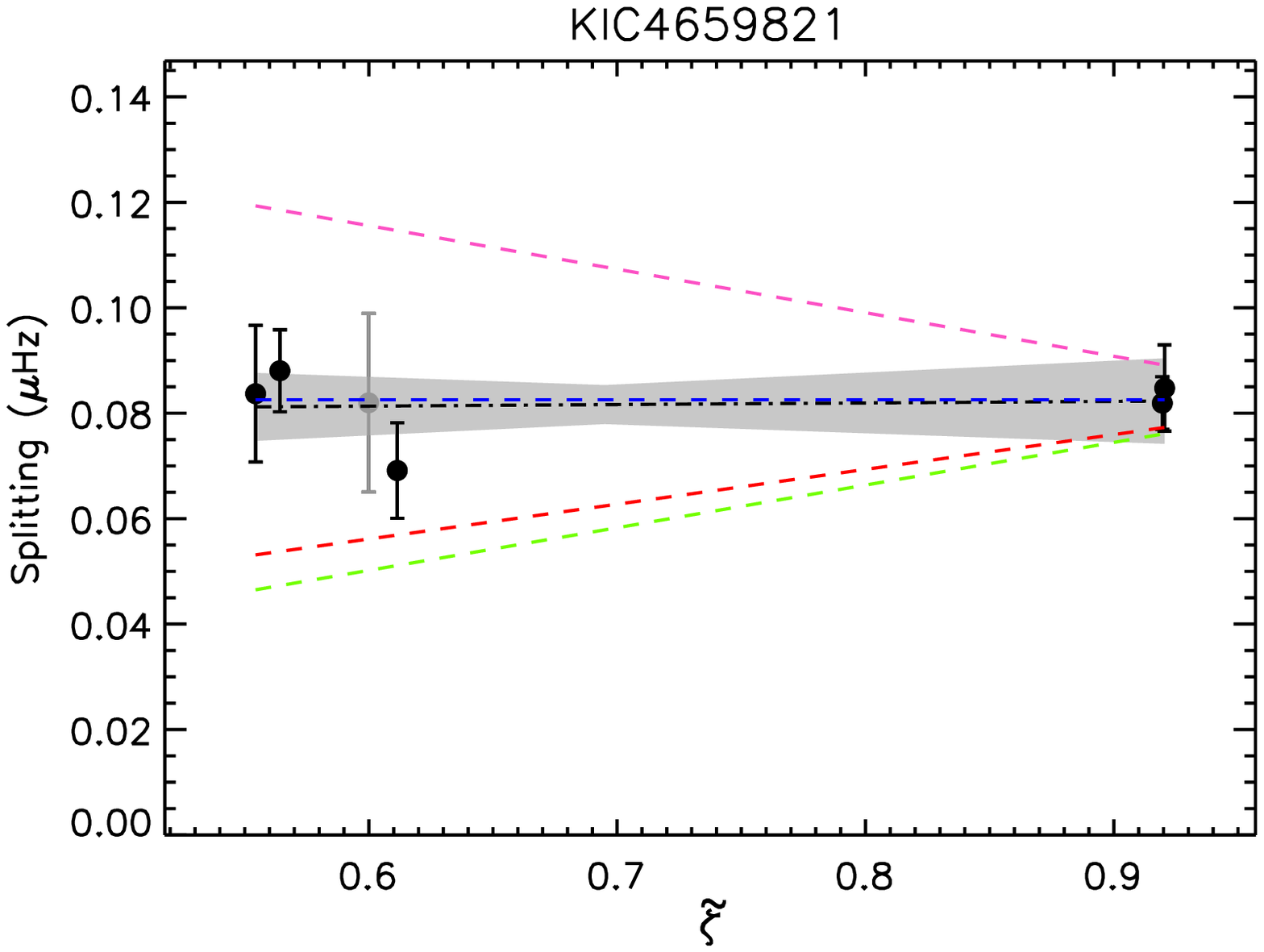}
\includegraphics[width=7.8cm]{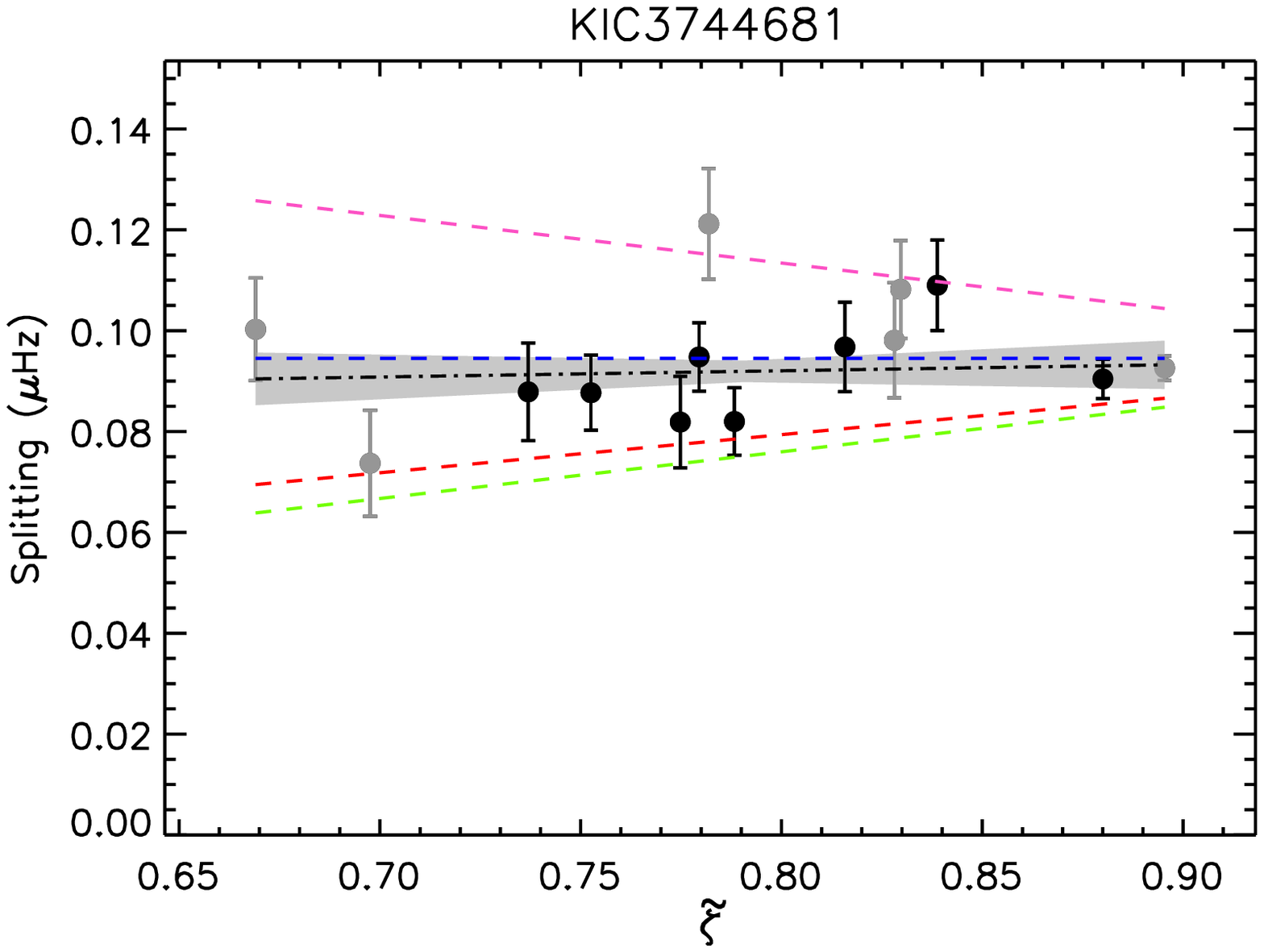}
\includegraphics[width=7.8cm]{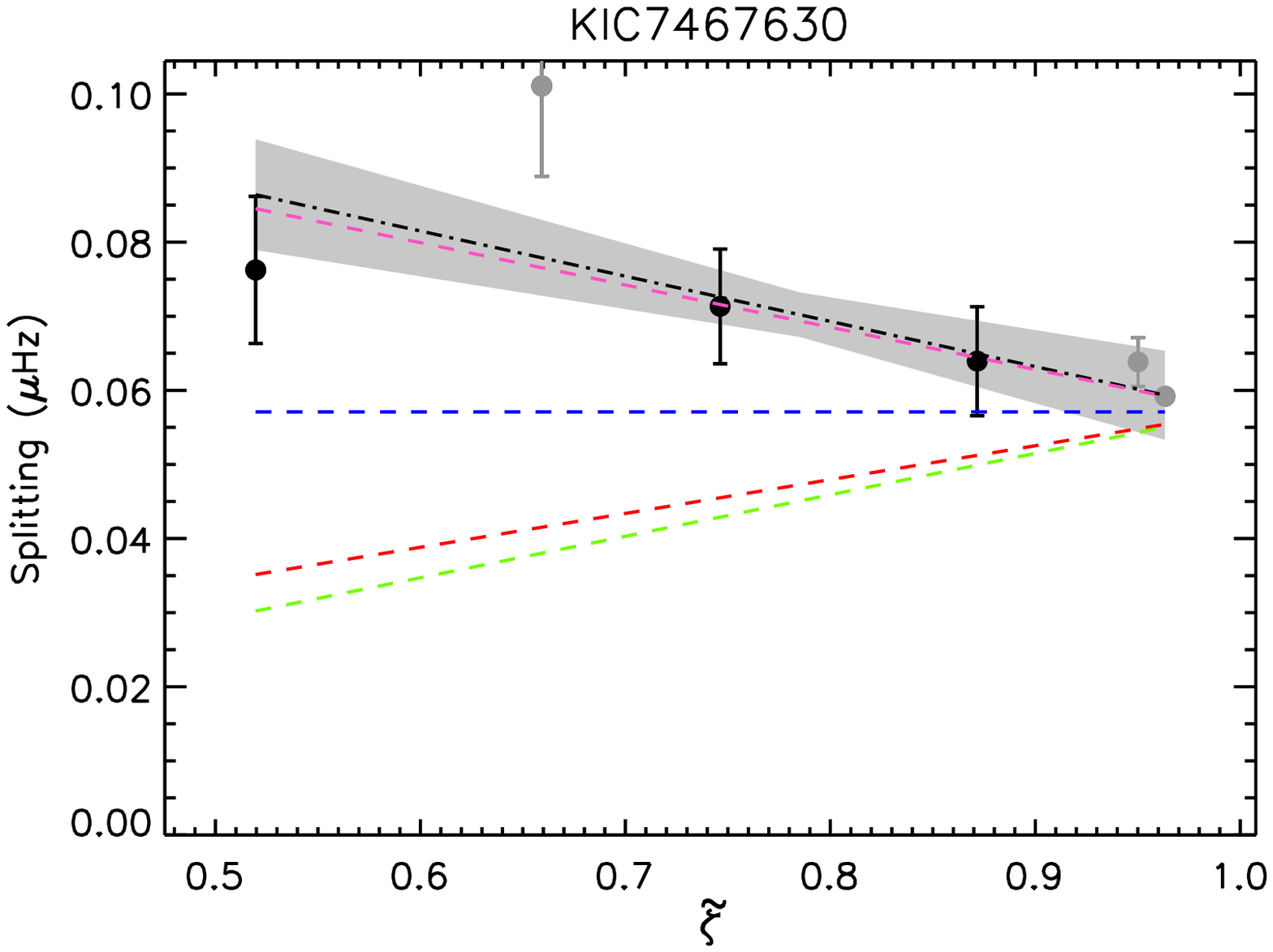}
\includegraphics[width=7.8cm]{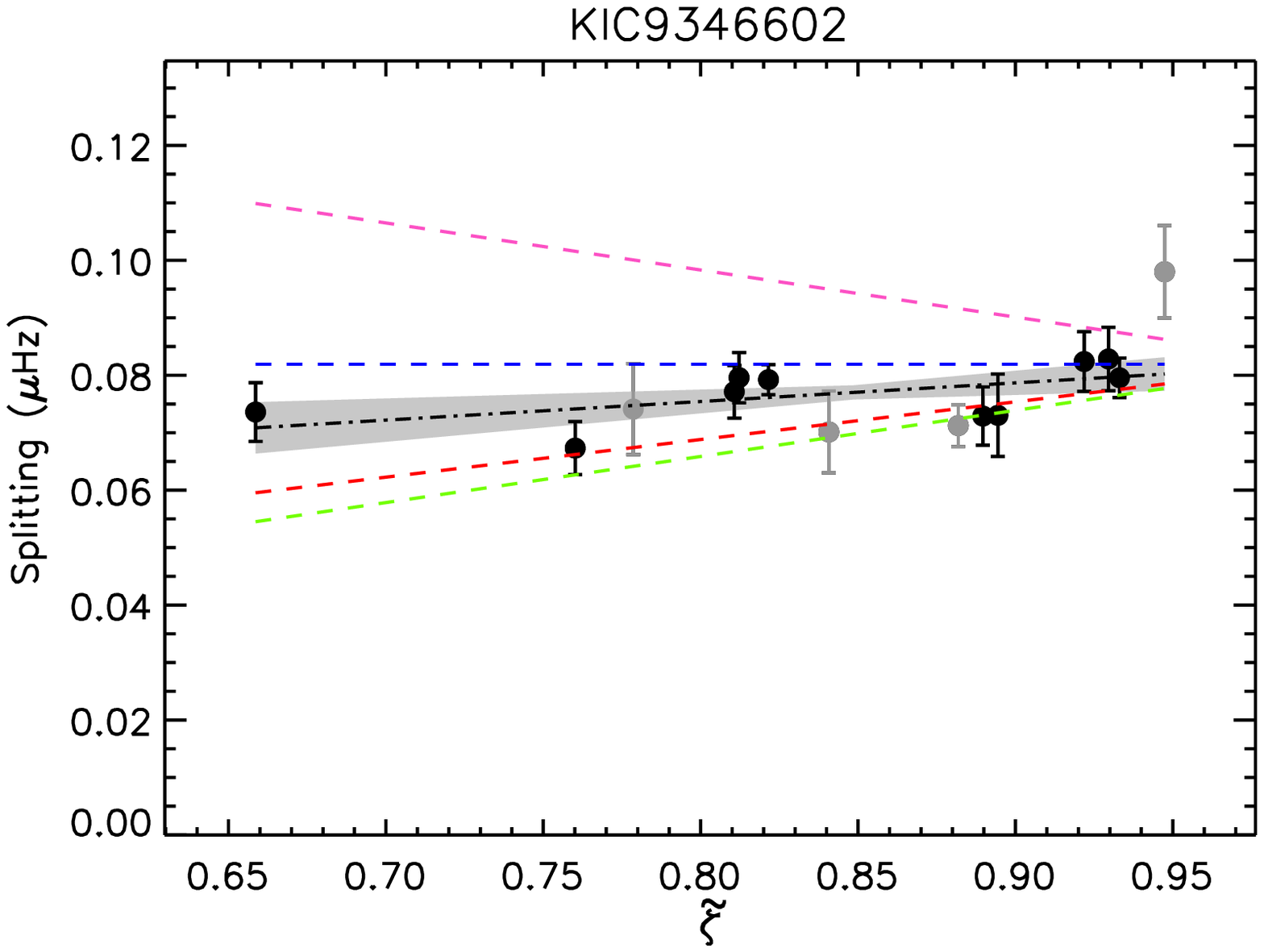}
\includegraphics[width=7.8cm]{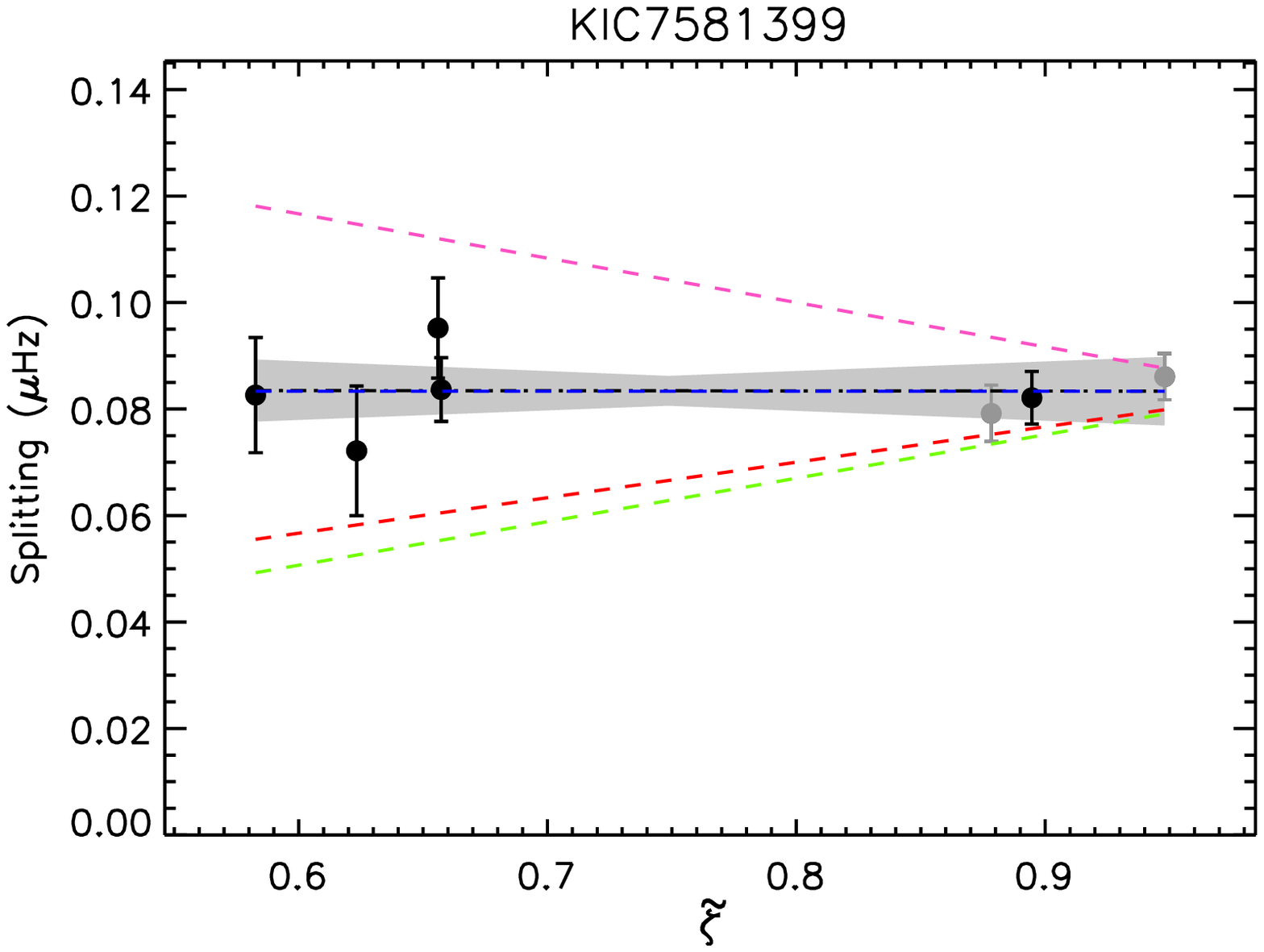}
\end{center}
\caption{Rotational splittings as a function of the parameter $\tilde{\zeta}$ which is an estimate of the mode trapping (see text).
The circles correspond to the observed splittings (black symbols indicate modes that passed both frequentist and Bayesian significance tests and gray symbols modes that passed the frequentist test only). The black dot-dashed line shows a linear regression of the observed splittings as a function of $\tilde{\zeta}$ and the gray shaded area indicates the uncertainties on the regression parameters. The dashed lines indicate theoretical splittings for 2-zone models that reproduce the observed rotation in the g-mode cavity and with $\Omega_{\rm core}/\Omega_{\rm surf}$ of 1, 2, 10, or 100 from top to bottom (same colors as in Fig. \ref{fig_zeta_split}).
\label{fig_zeta_split_others}}
\end{figure*}

In principle, the rotational kernels of the modes are required to extract information about the rotation profile from the mode splittings. However, G13 showed that the rotational splittings can be related to the parameter $\zeta$ through a linear dependence whose coefficients depend only on the mean rotation rate in the g-mode cavity $\langle\Omega_{\rm g}\rangle$ and  the mean rotation rate in the p-mode cavity $\langle\Omega_{\rm p}\rangle$:
\begin{linenomath*}
\begin{equation}
\delta\nu_{\rm s} = \left( \frac{\langle\Omega_{\rm g}\rangle/(2\pi)}{2} - \langle\Omega_{\rm p}\rangle/(2\pi) \right) \zeta + \langle\Omega_{\rm p}\rangle/(2\pi)
\label{eq_ompg}
\end{equation}
\end{linenomath*}
Estimates of $\langle\Omega_{\rm g}\rangle$ and $\langle\Omega_{\rm p}\rangle$ can thus be obtained by performing a linear regression of the function $\delta\nu(\zeta)$. This was already applied to early red giants by D14, who found that it provided results in excellent agreement with the inversions performed from stellar models. However, they had been using the trapping parameters $\zeta$ computed from stellar models. We here tested the method when replacing the parameters $\zeta$ by their approximate expression $\tilde{\zeta}$ computed with Eq. \ref{eq_zetatilde}. For \cible, such an approach yielded 
\begin{linenomath*}
\begin{align}
\langle\Omega_{\rm g}\rangle/(2\pi) & = 164\pm14\;\hbox{nHz} \nonumber \\
\langle\Omega_{\rm p}\rangle/(2\pi) & = 88\pm19\;\hbox{nHz} \nonumber \\
\langle\Omega_{\rm g}\rangle/\langle\Omega_{\rm p}\rangle & = 1.9\pm0.4 \nonumber
\end{align}
\end{linenomath*}
These results are in very good agreement with the values obtained the mode eigenfunctions of the best-fit model in Sect. \ref{sect_inversions}. This validates the simplified approach proposed by G13. 

\subsection{Constraints on the core-envelope contrast for seven \kepler\ secondary clump star \label{sect_other_stars}}

To estimate the amount of differential rotation for the seven targets selected in our sample, we have used the simplified approach of G13 that was described and validated with the test-case of \cible\ in Sect. \ref{sect_G13}. For each star, the trapping of the mode was estimated by computing the parameter $\tilde{\zeta}$ using Eq. \ref{eq_zetatilde}. Fig. \ref{fig_zeta_split_others} shows the mode splittings that passed the statistical test described in Sect. \ref{sect_analysis} as a function of $\tilde{\zeta}$. As was already observed for \cible\ in the previous section, it is striking to see how little the splittings vary with the mode trapping. For one star of the sample (KIC7467630), the splittings of p-dominated modes are found to be larger than the splittings of g-dominated modes.

We then used Eq. \ref{eq_ompg} to obtain estimates of $\langle\Omega_{\rm g}\rangle$ and $\langle\Omega_{\rm p}\rangle$ for all the stars of the sample. The results are given in Table \ref{tab_other_stars}. The ratios $\langle\Omega_{\rm g}\rangle/\langle\Omega_{\rm p}\rangle$ were found to vary from 1.3 to 3.2 for all the targets. This clearly confirms that the radial differential rotation in secondary clump stars is much weaker than it is for RGB stars, as was suggested by the analysis of \cible. However, we can exclude the scenario of a SB rotation for six of the seven selected targets, the last one (KIC7467630) being marginally consistent with a SB-rotating profile.


\begin{table*}
\caption{Spectroscopic results for the three targets that were observed with the HERMES spectrograph \label{tab_spec_values}}
\begin{center}
\begin{tabular}{l c c | c c c}
\hline\hline
\T KIC & RV& LB$_{\rm stellar}\,\sin i$ & inclination $i$  & LB$_{\rm stellar}$ & upper limit on $\Omega_{\rm surf}/(2\pi)$ \\
\B & (\kms) & (\kms) & ($^\circ$) & (\kms) & (nHz) \\
\hline
\T 4659821	& $-35.32\pm0.01$ & $8.5\pm0.5$  & $79\pm4$ & $8.6\pm0.6$ & $232\pm25$ \\
9346602	& $-11.90\pm0.01$ & $8.4\pm0.5$  & $83\pm2$ & $8.5\pm0.5$ & $158\pm21$ \\
\B 7581399	& $-23.81\pm0.01$ & $8.9\pm0.7$ & $79\pm5$ & $9.1\pm0.9$ & $193\pm29$ \\
\hline
\end{tabular}
\end{center}
\end{table*}

\section{Spectroscopic observations \label{sect_spectro}}

To complement seismic estimates of the rotation profile, we tried to obtain constraints on the surface rotation rate using high-resolution spectroscopic observations. For this purpose, we have observed three stars of the sample with the {High Efficiency and Resolution Mercator \'Echelle Spectrograph}  (HERMES, \citealt{raskin11}, \citealt{raskin14}) mounted on the 1.2\,m Mercator Telescope at the Spanish Observatorio del Roque de los Muchachos of the Instituto de Astrof\'{\i}sica de Canarias, with a resolving power of R$_{\rm HERMES}$\,=\,$\lambda$/$\Delta\lambda$$\,\simeq$\;85000, where $\lambda$ and $\Delta\lambda$ are the wavelength and the width per wavelength bin, respectively. The raw spectra were reduced and wavelength calibrated through Thorium-Argon reference spectra with the current version of the HERMES data reduction pipeline (Version 5, \citealt{raskin11}). The radial velocities (RV, Table\,\ref{tab_spec_values}) were obtained through a weighted cross-correlation of the wavelength range between 478 and 653 nm of each spectrum with a discrete G2 template (\citealt{raskin11}, \citealt{raskin14}). The post-processing and renormalization of the spectra were performed following \cite{beck15}.

The total line broadening $LB$ was determined from the equivalent width of seven unblended Fe\textsc{I} lines, following the procedure of \cite{gray05} and \cite{hekker07}. The average broadening and its uncertainty, reported in Table\,\ref{tab_spec_values} for the three stars were derived from the average and standard deviation of the individual total line broadening of the seven unblended FeI lines and deconvolved with the instrumental broadening of 1.76\,\kms. The stellar line broadening consists in contributions of the projected surface velocity, $v\sin i$ and the macro turbulence, $v_{\rm macro}$ such that
\begin{equation}
\hbox{LB}_{\rm stellar} \, \sin i = \sqrt{(v\sin i)^2 + v_{\rm macro}^2}.
\end{equation}
Disentangling the two contributions is challenging for red giant stars, as they have very narrow lines and $v_{\rm macro}$ was shown to have substantial contributions for this type of stars. Following the scaling relations from \cite{hekker07} for the value of $v_{\rm macro}$, we estimated the contributions of both parameters to be on the order of $\sim$~6\,\kms. However, owing to the large scatter in the data from which the scaling relations were obtained, we refrain from deriving a $v\sin i$ for each star. Instead, we use the stellar line broadening as an upper limit to the value of $v\sin i$ by assuming that $v_{\rm macro}=0$. 

In the present case, the inclination angles $i$ of the stars can be estimated from seismology. For each star, the extraction of each rotational splitting provided an estimate of $i$ (see Tables \ref{tab_lkhd} and \ref{tab_lkhd2}). By bringing these estimates together, we obtained estimates of $i$ for the three stars, which are listed in Table \ref{tab_spec_values}. We used these values to correct the stellar line broadening from the inclination effect, and thus obtained upper limits to the surface velocity of these stars. Since estimates of the radius of the three stars were already obtained from seismic scalings (see Table \ref{tab_targets}), we could translate these velocities into upper limits for the surface rotation rates, which are given in the last column of Table \ref{tab_spec_values}. These upper limits are of the order of the core rotation rates found through seismology (see Table \ref{tab_other_stars}), so the spectroscopic observations can be used only to rule out that the surface might be rotating faster than the core, but unfortunately not to test the core-envelope ratio that was obtained from seismology in Sect. \ref{sect_interpretation}.

\section{Discussion and conclusion \label{sect_conclusion}}

The aim of this paper was to measure the amount of differential rotation in secondary clump stars (intermediate-mass core helium burning stars) from seismology. This required to measure the rotational splittings of both g-dominated and p-dominated modes from the oscillation spectra of these stars. This was a challenge because the splittings of clump stars are comparable to the mode widths, particularly for p-dominated modes whose lifetimes are shorter. We followed both a frequentist approach using a maximum likelihood (MLE) method and a Bayesian approach with a Markov chain Monte-Carlo method to select only the statistically reliable splittings. We found a sample of seven \kepler\ secondary clump stars with both g and p-dominated reliable splittings.

To measure the amount of differential rotation in the selected stars using the mode splittings, we used an approach based on asymptotic theory proposed by G13, which has the advantage of being model-independent. We first validated this method on secondary clump stars by confronting it to the results of classical inversion techniques for one target of the sample (KIC7581399), and we proposed a slight modification to it. We applied this method to the selected stars and found evidence for a weak differential rotation in six out of the seven stars of the sample, with ratios between the core and envelope rotation rates ranging from $1.8\pm0.3$ and $3.2\pm1.0$. The last target was found to be marginally consistent with a solid-body (SB) rotation profile. We tried to complement the seismic measurement of the internal rotation with spectroscopic measurements of the surface velocity. High-resolution spectra were obtained for three of the selected targets, from which we deduced upper limits to the surface rotation rate. These limits confirmed that the surface cannot spin faster than the core, but they were not sufficient to test the amount of differential rotation found with seismology.

Our results clearly show that secondary clump stars have a much milder differential rotation than their RGB counterparts, for which core-envelope rotation ratios above 20 were observed (e.g. G13). Following the approach of \cite{tayar13}, we compared the total AM of the secondary clump stars of our sample to the estimated AM at the terminal-age main sequence (TAMS). Assuming SB rotation at the TAMS and typical surface rotation rates (50 to 150 \kms), we expect\footnote{To derive this estimate, we used the moment of inertia of the optimal model derived for KIC7581399 (Sect. \ref{sect_model}) whose evolution was stopped at the TAMS.} a total AM at the end of the MS around $(4\pm2)$ $10^{49}$ erg.s. By comparison, using the mean core and envelope rotation rates derived in this study and our optimal model of KIC7581399 as a reference model for secondary clump stars, we obtained estimates of the total AM on the clump around 2 $10^{49}$ erg.s for the seven stars of the sample. This value is of the same order of magnitude as the total AM expected at the TAMS, and the slightly lower AM found on the clump might be caused by mass loss, which is expected to occur between the TAMS and the triggering of core-He burning (not taken into account by \citealt{tayar13}).


The weak radial differential rotation that we find for secondary clump stars means that a very efficient redistribution of AM occurs either before or right after these stars settle on the clump. If this strong coupling occurs before He-burning is triggered in the core, it has to operate on a very short timescale since these stars cross the subgiant phase on a thermal timescale. To illustrate this point, a 3-$M_\odot$ star evolves from the MS turnoff to the clump in only 15 Myr, which amounts to about 5\% of the time it spends on the MS. For comparison, it takes around 700 Myr for a 1.5-$M_\odot$ star, i.e. about 34\% of the time spent on the MS. So if the same mechanism is responsible for AM redistribution in low-mass and intermediate-mass stars, the present results provide more stringent constraints on its timescale. The other possible explanation is that AM could be redistributed after the triggering of He-core burning. Internal gravity waves (IGW) excited both at the bottom of the convective envelope and at the top of the convective core generated by He-burning could efficiently couple the core to the envelope (\citealt{talon08}, \citealt{fuller14}). \cite{maeder14} have also recently shown from order-of-magnitude estimates that a fossil magnetic field attached to the core could produce efficient coupling between the core and the envelope during the He-burning phase. To determine at which stage the coupling occurs would require to find intermediate-mass stars in the subgiant phase or at the bottom of the RGB. Even though this phase is very fast, there might be a handful of such targets among the \kepler\ targets. 

Regardless of when the AM redistribution occurs, it has to happen on a very short timescale, which should provide valuable observational constraints to theoretical models of AM transport mechanisms in intermediate-mass stars. It is important to note that even though an efficient AM transport is needed, it does not generally lead to SB rotation since the SB rotation case was significantly ruled out for six of the seven targets of our sample. Interestingly, the two targets that were found to have the lowest core-envelope rotation contrast are the ones with the largest values of $\Delta\Pi_1$. Since $\Delta\Pi_1$ increases during the core He-burning phase\footnote{the increase in $\Delta\Pi_1$ is due to the increasing core mass as the H-burning shell progresses outwards, which induced a decrease in the \vaisala\ frequency} (see e.g. \citealt{mosser14}), this could mean that SB rotation progressively builds up during this period. More data are required to confirm this hypothesis. In this context, the observations of the selected space mission Plato (\citealt{plato}) will be particularly useful.


\begin{acknowledgements}
S.D. acknowledges support from the PNPS under the grant "Rotation interne et magn\'etisme des sous-g\'eantes et g\'eantes Kepler" and from the Centre National d'\'Etudes Spatiales (CNES). J.B. thanks O. Benomar for fruitful discussions about automated parallel tempering MCMC. P.G.B., B.M., R.A.G., and M.J.G. acknowledge the ANR (Agence Nationale de la Recherche, France) program IDEE (n$^\circ$ ANR-12-BS05-0008) "Int\'eraction Des \'Etoiles et des Exoplan\`etes". The ground-based observations are based on spectroscopy made with the Mercator Telescope, operated on the island of La Palma by the Flemish Community, at the Spanish Observatorio del Roque de los Muchachos of the Instituto de Astrof\'{\i}sica de Canarias.
\end{acknowledgements}

\bibliographystyle{aa.bst} 
\bibliography{biblio} 

\begin{thebibliography}{65}
\expandafter\ifx\csname natexlab\endcsname\relax\def\natexlab#1{#1}\fi

\bibitem[{{Aerts} {et~al.}(2010){Aerts}, {Christensen-Dalsgaard}, \&
  {Kurtz}}]{aerts10}
{Aerts}, C., {Christensen-Dalsgaard}, J., \& {Kurtz}, D.~W. 2010,
  {Asteroseismology}

\bibitem[{{Angulo} {et~al.}(1999){Angulo}, {Arnould}, {Rayet}, {Descouvemont},
  {Baye}, {Leclercq-Willain}, {Coc}, {Barhoumi}, {Aguer}, {Rolfs}, {Kunz},
  {Hammer}, {Mayer}, {Paradellis}, {Kossionides}, {Chronidou}, {Spyrou},
  {degl'Innocenti}, {Fiorentini}, {Ricci}, {Zavatarelli}, {Providencia},
  {Wolters}, {Soares}, {Grama}, {Rahighi}, {Shotter}, \& {Lamehi
  Rachti}}]{angulo99}
{Angulo}, C., {Arnould}, M., {Rayet}, M., {et~al.} 1999, Nuclear Physics A,
  656, 3

\bibitem[{{Appourchaux} {et~al.}(1998){Appourchaux}, {Gizon}, \&
  {Rabello-Soares}}]{appourchaux98}
{Appourchaux}, T., {Gizon}, L., \& {Rabello-Soares}, M. 1998, \aaps, 132, 107

\bibitem[{{Ballot} {et~al.}(2006){Ballot}, {Garc{\'{\i}}a}, \&
  {Lambert}}]{ballot06}
{Ballot}, J., {Garc{\'{\i}}a}, R.~A., \& {Lambert}, P. 2006, \mnras, 369, 1281

\bibitem[{{Beck} {et~al.}(2014){Beck}, {Hambleton}, {Vos}, {Kallinger},
  {Bloemen}, {Tkachenko}, {Garc{\'{\i}}a}, {{\O}stensen}, {Aerts}, {Kurtz}, {De
  Ridder}, {Hekker}, {Pavlovski}, {Mathur}, {De Smedt}, {Derekas}, {Corsaro},
  {Mosser}, {Van Winckel}, {Huber}, {Degroote}, {Davies}, {Pr{\v s}a},
  {Debosscher}, {Elsworth}, {Nemeth}, {Siess}, {Schmid}, {P{\'a}pics}, {de
  Vries}, {van Marle}, {Marcos-Arenal}, \& {Lobel}}]{beck14}
{Beck}, P.~G., {Hambleton}, K., {Vos}, J., {et~al.} 2014, \aap, 564, A36

\bibitem[{{Beck} {et~al.}(2012){Beck}, {Montalban}, {Kallinger}, {De Ridder},
  {Aerts}, {Garc{\'{\i}}a}, {Hekker}, {Dupret}, {Mosser}, {Eggenberger},
  {Stello}, {Elsworth}, {Frandsen}, {Carrier}, {Hillen}, {Gruberbauer},
  {Christensen-Dalsgaard}, {Miglio}, {Valentini}, {Bedding}, {Kjeldsen},
  {Girouard}, {Hall}, \& {Ibrahim}}]{beck12}
{Beck}, P.~G., {Montalban}, J., {Kallinger}, T., {et~al.} 2012, \nat, 481, 55

\bibitem[{{Beck} {et~al.}(2015){Beck}, {Van Reeth}, {Tachenko}, {Allende
  Prieto}, {Raskin}, {Van Winckel}, {Gar{\' i}a}, {G.}, \& {G.}}]{beck15}
{Beck}, P.~G., {Van Reeth}, T., {Tachenko}, A., {et~al.} 2015, A\&A submitted

\bibitem[{{Bedding} {et~al.}(2011){Bedding}, {Mosser}, {Huber},
  {Montalb{\'a}n}, {Beck}, {Christensen-Dalsgaard}, {Elsworth},
  {Garc{\'{\i}}a}, {Miglio}, {Stello}, {White}, {De Ridder}, {Hekker}, {Aerts},
  {Barban}, {Belkacem}, {Broomhall}, {Brown}, {Buzasi}, {Carrier}, {Chaplin},
  {di Mauro}, {Dupret}, {Frandsen}, {Gilliland}, {Goupil}, {Jenkins},
  {Kallinger}, {Kawaler}, {Kjeldsen}, {Mathur}, {Noels}, {Aguirre}, \&
  {Ventura}}]{bedding11}
{Bedding}, T.~R., {Mosser}, B., {Huber}, D., {et~al.} 2011, \nat, 471, 608

\bibitem[{{Belkacem} {et~al.}(2011){Belkacem}, {Goupil}, {Dupret}, {Samadi},
  {Baudin}, {Noels}, \& {Mosser}}]{belkacem11}
{Belkacem}, K., {Goupil}, M.~J., {Dupret}, M.~A., {et~al.} 2011, \aap, 530,
  A142

\bibitem[{{Benomar} {et~al.}(2009){Benomar}, {Appourchaux}, \&
  {Baudin}}]{benomar09a}
{Benomar}, O., {Appourchaux}, T., \& {Baudin}, F. 2009, \aap, 506, 15

\bibitem[{{B{\"o}hm-Vitense}(1958)}]{bohm58}
{B{\"o}hm-Vitense}, E. 1958, \zap, 46, 108

\bibitem[{{Brown} {et~al.}(1991){Brown}, {Gilliland}, {Noyes}, \&
  {Ramsey}}]{brown91}
{Brown}, T.~M., {Gilliland}, R.~L., {Noyes}, R.~W., \& {Ramsey}, L.~W. 1991,
  \apj, 368, 599

\bibitem[{{Cantiello} {et~al.}(2014){Cantiello}, {Mankovich}, {Bildsten},
  {Christensen-Dalsgaard}, \& {Paxton}}]{cantiello14}
{Cantiello}, M., {Mankovich}, C., {Bildsten}, L., {Christensen-Dalsgaard}, J.,
  \& {Paxton}, B. 2014, \apj, 788, 93

\bibitem[{{Ceillier} {et~al.}(2013){Ceillier}, {Eggenberger}, {Garc{\'{\i}}a},
  \& {Mathis}}]{ceillier13}
{Ceillier}, T., {Eggenberger}, P., {Garc{\'{\i}}a}, R.~A., \& {Mathis}, S.
  2013, \aap, 555, A54

\bibitem[{{Chaplin} {et~al.}(1999){Chaplin}, {Christensen-Dalsgaard},
  {Elsworth}, {Howe}, {Isaak}, {Larsen}, {New}, {Schou}, {Thompson}, \&
  {Tomczyk}}]{chaplin99}
{Chaplin}, W.~J., {Christensen-Dalsgaard}, J., {Elsworth}, Y., {et~al.} 1999,
  \mnras, 308, 405

\bibitem[{{Charbonnel} \& {Talon}(2005)}]{charbonnel05}
{Charbonnel}, C. \& {Talon}, S. 2005, Science, 309, 2189

\bibitem[{{Deheuvels} {et~al.}(2014){Deheuvels}, {Do{\u g}an}, {Goupil},
  {Appourchaux}, {Benomar}, {Bruntt}, {Campante}, {Casagrande}, {Ceillier},
  {Davies}, {De Cat}, {Fu}, {Garc{\'{\i}}a}, {Lobel}, {Mosser}, {Reese},
  {Regulo}, {Schou}, {Stahn}, {Thygesen}, {Yang}, {Chaplin},
  {Christensen-Dalsgaard}, {Eggenberger}, {Gizon}, {Mathis},
  {Molenda-{\.Z}akowicz}, \& {Pinsonneault}}]{deheuvels14}
{Deheuvels}, S., {Do{\u g}an}, G., {Goupil}, M.~J., {et~al.} 2014, \aap, 564,
  A27

\bibitem[{{Deheuvels} {et~al.}(2012){Deheuvels}, {Garc{\'{\i}}a}, {Chaplin},
  {Basu}, {Antia}, {Appourchaux}, {Benomar}, {Davies}, {Elsworth}, {Gizon},
  {Goupil}, {Reese}, {Regulo}, {Schou}, {Stahn}, {Casagrande},
  {Christensen-Dalsgaard}, {Fischer}, {Hekker}, {Kjeldsen}, {Mathur}, {Mosser},
  {Pinsonneault}, {Valenti}, {Christiansen}, {Kinemuchi}, \&
  {Mullally}}]{deheuvels12}
{Deheuvels}, S., {Garc{\'{\i}}a}, R.~A., {Chaplin}, W.~J., {et~al.} 2012, \apj,
  756, 19

\bibitem[{{Deheuvels} \& {Michel}(2011)}]{deheuvels11}
{Deheuvels}, S. \& {Michel}, E. 2011, \aap, 535, A91

\bibitem[{{Eggenberger} {et~al.}(2012){Eggenberger}, {Montalb{\'a}n}, \&
  {Miglio}}]{eggenberger12}
{Eggenberger}, P., {Montalb{\'a}n}, J., \& {Miglio}, A. 2012, \aap, 544, L4

\bibitem[{{Fuller} {et~al.}(2014){Fuller}, {Lecoanet}, {Cantiello}, \&
  {Brown}}]{fuller14}
{Fuller}, J., {Lecoanet}, D., {Cantiello}, M., \& {Brown}, B. 2014, \apj, 796,
  17

\bibitem[{{Garc{\'{\i}}a} {et~al.}(2011){Garc{\'{\i}}a}, {Hekker}, {Stello},
  {Guti{\'e}rrez-Soto}, {Handberg}, {Huber}, {Karoff}, {Uytterhoeven},
  {Appourchaux}, {Chaplin}, {Elsworth}, {Mathur}, {Ballot},
  {Christensen-Dalsgaard}, {Gilliland}, {Houdek}, {Jenkins}, {Kjeldsen},
  {McCauliff}, {Metcalfe}, {Middour}, {Molenda-Zakowicz}, {Monteiro}, {Smith},
  \& {Thompson}}]{garcia11}
{Garc{\'{\i}}a}, R.~A., {Hekker}, S., {Stello}, D., {et~al.} 2011, \mnras, 414,
  L6

\bibitem[{{Garc{\'{\i}}a} {et~al.}(2008){Garc{\'{\i}}a}, {Mathur}, {Ballot},
  {Eff-Darwich}, {Jim{\'e}nez-Reyes}, \& {Korzennik}}]{garcia08}
{Garc{\'{\i}}a}, R.~A., {Mathur}, S., {Ballot}, J., {et~al.} 2008, \solphys,
  251, 119

\bibitem[{{Gizon} \& {Solanki}(2003)}]{gizon03}
{Gizon}, L. \& {Solanki}, S.~K. 2003, \apj, 589, 1009

\bibitem[{{Gough} \& {McIntyre}(1998)}]{gough98}
{Gough}, D.~O. \& {McIntyre}, M.~E. 1998, \nat, 394, 755

\bibitem[{{Goupil} {et~al.}(2013){Goupil}, {Mosser}, {Marques}, {Ouazzani},
  {Belkacem}, {Lebreton}, \& {Samadi}}]{goupil13}
{Goupil}, M.~J., {Mosser}, B., {Marques}, J.~P., {et~al.} 2013, \aap, 549, A75

\bibitem[{{Gray}(2005)}]{gray05}
{Gray}, D.~F. 2005, {The Observation and Analysis of Stellar Photospheres}

\bibitem[{{Grevesse} \& {Noels}(1993)}]{grevesse93}
{Grevesse}, N. \& {Noels}, A. 1993, in Origin and Evolution of the Elements,
  ed. {N.~Prantzos, E.~Vangioni-Flam, \& M.~Casse}, 15--25

\bibitem[{{Harvey}(1985)}]{harvey85}
{Harvey}, J. 1985, {High-resolution helioseismology}, Tech. rep.

\bibitem[{{Hekker} \& {Mel{\'e}ndez}(2007)}]{hekker07}
{Hekker}, S. \& {Mel{\'e}ndez}, J. 2007, A\&A, 475, 1003

\bibitem[{{Jenkins} {et~al.}(2010){Jenkins}, {Caldwell}, {Chandrasekaran},
  {Twicken}, {Bryson}, {Quintana}, {Clarke}, {Li}, {Allen}, {Tenenbaum}, {Wu},
  {Klaus}, {Van Cleve}, {Dotson}, {Haas}, {Gilliland}, {Koch}, \&
  {Borucki}}]{jenkins10}
{Jenkins}, J.~M., {Caldwell}, D.~A., {Chandrasekaran}, H., {et~al.} 2010,
  \apjl, 713, L120

\bibitem[{{Kallinger} {et~al.}(2014){Kallinger}, {De Ridder}, {Hekker},
  {Mathur}, {Mosser}, {Gruberbauer}, {Garc{\'{\i}}a}, {Karoff}, \&
  {Ballot}}]{kallinger14}
{Kallinger}, T., {De Ridder}, J., {Hekker}, S., {et~al.} 2014, \aap, 570, A41

\bibitem[{{Karoff} {et~al.}(2013){Karoff}, {Campante}, {Ballot}, {Kallinger},
  {Gruberbauer}, {Garc{\'{\i}}a}, {Caldwell}, {Christiansen}, \&
  {Kinemuchi}}]{karoff13}
{Karoff}, C., {Campante}, T.~L., {Ballot}, J., {et~al.} 2013, \apj, 767, 34

\bibitem[{{Kjeldsen} {et~al.}(2008){Kjeldsen}, {Bedding}, \&
  {Christensen-Dalsgaard}}]{kjeldsen08}
{Kjeldsen}, H., {Bedding}, T.~R., \& {Christensen-Dalsgaard}, J. 2008, \apjl,
  683, L175

\bibitem[{{Kurtz} {et~al.}(2014){Kurtz}, {Saio}, {Takata}, {Shibahashi},
  {Murphy}, \& {Sekii}}]{kurtz14}
{Kurtz}, D.~W., {Saio}, H., {Takata}, M., {et~al.} 2014, \mnras, 444, 102

\bibitem[{{Lagarde} {et~al.}(2012){Lagarde}, {Decressin}, {Charbonnel},
  {Eggenberger}, {Ekstr{\"o}m}, \& {Palacios}}]{lagarde12}
{Lagarde}, N., {Decressin}, T., {Charbonnel}, C., {et~al.} 2012, \aap, 543,
  A108

\bibitem[{{Lomb}(1976)}]{lomb76}
{Lomb}, N.~R. 1976, \apss, 39, 447

\bibitem[{{Maeder} \& {Meynet}(2014)}]{maeder14}
{Maeder}, A. \& {Meynet}, G. 2014, \apj, 793, 123

\bibitem[{{Marques} {et~al.}(2013){Marques}, {Goupil}, {Lebreton}, {Talon},
  {Palacios}, {Belkacem}, {Ouazzani}, {Mosser}, {Moya}, {Morel}, {Pichon},
  {Mathis}, {Zahn}, {Turck-Chi{\`e}ze}, \& {Nghiem}}]{marques13}
{Marques}, J.~P., {Goupil}, M.~J., {Lebreton}, Y., {et~al.} 2013, \aap, 549,
  A74

\bibitem[{{Mathis} \& {Zahn}(2004)}]{mathis04}
{Mathis}, S. \& {Zahn}, J. 2004, \aap, 425, 229

\bibitem[{{Montalb{\'a}n} {et~al.}(2013){Montalb{\'a}n}, {Miglio}, {Noels},
  {Dupret}, {Scuflaire}, \& {Ventura}}]{montalban13}
{Montalb{\'a}n}, J., {Miglio}, A., {Noels}, A., {et~al.} 2013, \apj, 766, 118

\bibitem[{{Mosser} {et~al.}(2014){Mosser}, {Benomar}, {Belkacem}, {Goupil},
  {Lagarde}, {Michel}, {Lebreton}, {Stello}, {Vrard}, {Barban}, {Bedding},
  {Deheuvels}, {Chaplin}, {De Ridder}, {Elsworth}, {Montalban}, {Noels},
  {Ouazzani}, {Samadi}, {White}, \& {Kjeldsen}}]{mosser14}
{Mosser}, B., {Benomar}, O., {Belkacem}, K., {et~al.} 2014, \aap, 572, L5

\bibitem[{{Mosser} {et~al.}(2012{\natexlab{a}}){Mosser}, {Goupil}, {Belkacem},
  {Marques}, {Beck}, {Bloemen}, {De Ridder}, {Barban}, {Deheuvels}, {Elsworth},
  {Hekker}, {Kallinger}, {Ouazzani}, {Pinsonneault}, {Samadi}, {Stello},
  {Garc{\'{\i}}a}, {Klaus}, {Li}, {Mathur}, \& {Morris}}]{mosser12b}
{Mosser}, B., {Goupil}, M.~J., {Belkacem}, K., {et~al.} 2012{\natexlab{a}},
  \aap, 548, A10

\bibitem[{{Mosser} {et~al.}(2012{\natexlab{b}}){Mosser}, {Goupil}, {Belkacem},
  {Michel}, {Stello}, {Marques}, {Elsworth}, {Barban}, {Beck}, {Bedding}, {De
  Ridder}, {Garc{\'{\i}}a}, {Hekker}, {Kallinger}, {Samadi}, {Stumpe},
  {Barclay}, \& {Burke}}]{mosser12a}
{Mosser}, B., {Goupil}, M.~J., {Belkacem}, K., {et~al.} 2012{\natexlab{b}},
  \aap, 540, A143

\bibitem[{{Mosser} {et~al.}(2013){Mosser}, {Michel}, {Belkacem}, {Goupil},
  {Baglin}, {Barban}, {Provost}, {Samadi}, {Auvergne}, \& {Catala}}]{mosser13}
{Mosser}, B., {Michel}, E., {Belkacem}, K., {et~al.} 2013, \aap, 550, A126

\bibitem[{{Paxton} {et~al.}(2011){Paxton}, {Bildsten}, {Dotter}, {Herwig},
  {Lesaffre}, \& {Timmes}}]{mesa}
{Paxton}, B., {Bildsten}, L., {Dotter}, A., {et~al.} 2011, \apjs, 192, 3

\bibitem[{{Pinsonneault} {et~al.}(2012){Pinsonneault}, {An},
  {Molenda-{\.Z}akowicz}, {Chaplin}, {Metcalfe}, \& {Bruntt}}]{pinsonneault12}
{Pinsonneault}, M.~H., {An}, D., {Molenda-{\.Z}akowicz}, J., {et~al.} 2012,
  \apjs, 199, 30

\bibitem[{{Raskin } \& {Van Winckel}(2014)}]{raskin14}
{Raskin }, G. \& {Van Winckel}, H. 2014, Astronomische Nachrichten, 335, 32

\bibitem[{{Raskin} {et~al.}(2011){Raskin}, {van Winckel}, {Hensberge},
  {Jorissen}, {Lehmann}, {Waelkens}, {Avila}, {de Cuyper}, {Degroote},
  {Dubosson}, {Dumortier}, {Fr{\'e}mat}, {Laux}, {Michaud}, {Morren}, {Perez
  Padilla}, {Pessemier}, {Prins}, {Smolders}, {van Eck}, \&
  {Winkler}}]{raskin11}
{Raskin}, G., {van Winckel}, H., {Hensberge}, H., {et~al.} 2011, \aap, 526, A69

\bibitem[{{Rauer} {et~al.}(2014){Rauer}, {Catala}, {Aerts}, {Appourchaux},
  {Benz}, {Brandeker}, {Christensen-Dalsgaard}, {Deleuil}, {Gizon}, {Goupil},
  {G{\"u}del}, {Janot-Pacheco}, {Mas-Hesse}, {Pagano}, {Piotto}, {Pollacco},
  {Santos}, {Smith}, {Su{\'a}rez}, {Szab{\'o}}, {Udry}, {Adibekyan}, {Alibert},
  {Almenara}, {Amaro-Seoane}, {Ammer-von Eiff}, {Asplund}, {Antonello},
  {Barnes}, {Baudin}, {Belkacem}, {Bergemann}, {Bihain}, {Birch}, {Bonfils},
  {Boisse}, {Bonomo}, {Borsa}, {Brand{\~a}o}, {Brocato}, {Brun}, {Burleigh},
  {Burston}, {Cabrera}, {Cassisi}, {Chaplin}, {Charpinet}, {Chiappini},
  {Church}, {Csizmadia}, {Cunha}, {Damasso}, {Davies}, {Deeg}, {D{\'{\i}}az},
  {Dreizler}, {Dreyer}, {Eggenberger}, {Ehrenreich}, {Eigm{\"u}ller},
  {Erikson}, {Farmer}, {Feltzing}, {Oliveira Fialho}, {Figueira}, {Forveille},
  {Fridlund}, {Garc{\'{\i}}a}, {Giommi}, {Giuffrida}, {Godolt}, {Gomes da
  Silva}, {Granzer}, {Grenfell}, {Grotsch-Noels}, {G{\"u}nther}, {Haswell},
  {Hatzes}, {H{\'e}brard}, {Hekker}, {Helled}, {Heng}, {Jenkins}, {Johansen},
  {Khodachenko}, {Kislyakova}, {Kley}, {Kolb}, {Krivova}, {Kupka}, {Lammer},
  {Lanza}, {Lebreton}, {Magrin}, {Marcos-Arenal}, {Marrese}, {Marques},
  {Martins}, {Mathis}, {Mathur}, {Messina}, {Miglio}, {Montalban}, {Montalto},
  {Monteiro}, {Moradi}, {Moravveji}, {Mordasini}, {Morel}, {Mortier},
  {Nascimbeni}, {Nelson}, {Nielsen}, {Noack}, {Norton}, {Ofir}, {Oshagh},
  {Ouazzani}, {P{\'a}pics}, {Parro}, {Petit}, {Plez}, {Poretti}, {Quirrenbach},
  {Ragazzoni}, {Raimondo}, {Rainer}, {Reese}, {Redmer}, {Reffert},
  {Rojas-Ayala}, {Roxburgh}, {Salmon}, {Santerne}, {Schneider}, {Schou},
  {Schuh}, {Schunker}, {Silva-Valio}, {Silvotti}, {Skillen}, {Snellen}, {Sohl},
  {Sousa}, {Sozzetti}, {Stello}, {Strassmeier}, {{\v S}vanda}, {Szab{\'o}},
  {Tkachenko}, {Valencia}, {Van Grootel}, {Vauclair}, {Ventura}, {Wagner},
  {Walton}, {Weingrill}, {Werner}, {Wheatley}, \& {Zwintz}}]{plato}
{Rauer}, H., {Catala}, C., {Aerts}, C., {et~al.} 2014, Experimental Astronomy

\bibitem[{{Rogers} \& {Nayfonov}(2002)}]{rogers02}
{Rogers}, F.~J. \& {Nayfonov}, A. 2002, \apj, 576, 1064

\bibitem[{{R{\"u}diger} {et~al.}(2014{\natexlab{a}}){R{\"u}diger}, {Gellert},
  {Spada}, \& {Tereshin}}]{rudiger14b}
{R{\"u}diger}, G., {Gellert}, M., {Spada}, F., \& {Tereshin}, I.
  2014{\natexlab{a}}, ArXiv e-prints

\bibitem[{{R{\"u}diger} {et~al.}(2014{\natexlab{b}}){R{\"u}diger}, {Schultz},
  \& {Kitchatinov}}]{rudiger14a}
{R{\"u}diger}, G., {Schultz}, M., \& {Kitchatinov}, L.~L. 2014{\natexlab{b}},
  ArXiv e-prints

\bibitem[{{Scargle}(1982)}]{scargle82}
{Scargle}, J.~D. 1982, \apj, 263, 835

\bibitem[{{Schou} {et~al.}(1998){Schou}, {Antia}, {Basu}, {Bogart}, {Bush},
  {Chitre}, {Christensen-Dalsgaard}, {di Mauro}, {Dziembowski}, {Eff-Darwich},
  {Gough}, {Haber}, {Hoeksema}, {Howe}, {Korzennik}, {Kosovichev}, {Larsen},
  {Pijpers}, {Scherrer}, {Sekii}, {Tarbell}, {Title}, {Thompson}, \&
  {Toomre}}]{schou98}
{Schou}, J., {Antia}, H.~M., {Basu}, S., {et~al.} 1998, \apj, 505, 390

\bibitem[{{Scuflaire} {et~al.}(2008){Scuflaire}, {Montalb{\'a}n}, {Th{\'e}ado},
  {Bourge}, {Miglio}, {Godart}, {Thoul}, \& {Noels}}]{losc}
{Scuflaire}, R., {Montalb{\'a}n}, J., {Th{\'e}ado}, S., {et~al.} 2008, \apss,
  316, 149

\bibitem[{{Spruit}(1999)}]{spruit99}
{Spruit}, H.~C. 1999, \aap, 349, 189

\bibitem[{{Spruit}(2002)}]{spruit02}
{Spruit}, H.~C. 2002, \aap, 381, 923

\bibitem[{{Stello} {et~al.}(2008){Stello}, {Bruntt}, {Preston}, \&
  {Buzasi}}]{stello08}
{Stello}, D., {Bruntt}, H., {Preston}, H., \& {Buzasi}, D. 2008, \apjl, 674,
  L53

\bibitem[{{Talon} \& {Charbonnel}(2008)}]{talon08}
{Talon}, S. \& {Charbonnel}, C. 2008, \aap, 482, 597

\bibitem[{{Tassoul}(1980)}]{tassoul80}
{Tassoul}, M. 1980, \apjs, 43, 469

\bibitem[{{Tayar} \& {Pinsonneault}(2013)}]{tayar13}
{Tayar}, J. \& {Pinsonneault}, M.~H. 2013, ArXiv e-prints

\bibitem[{{Unno} {et~al.}(1989){Unno}, {Osaki}, {Ando}, {Saio}, \&
  {Shibahashi}}]{unno89}
{Unno}, W., {Osaki}, Y., {Ando}, H., {Saio}, H., \& {Shibahashi}, H. 1989,
  {Nonradial oscillations of stars}, ed. {Unno, W., Osaki, Y., Ando, H., Saio,
  H., \& Shibahashi, H. (Tokyo: University of Tokyo Press).}

\bibitem[{{Wilks}(1938)}]{wilks38}
{Wilks}, S.~S. 1938, Ann. Math. Stat., 9, 60

\bibitem[{{Zahn}(1992)}]{zahn92}
{Zahn}, J.-P. 1992, \aap, 265, 115

\end{thebibliography}

\begin{appendix}

\section{Monte Carlo \label{app_wilks}}

\begin{figure}
\begin{center}
\includegraphics[width=9cm]{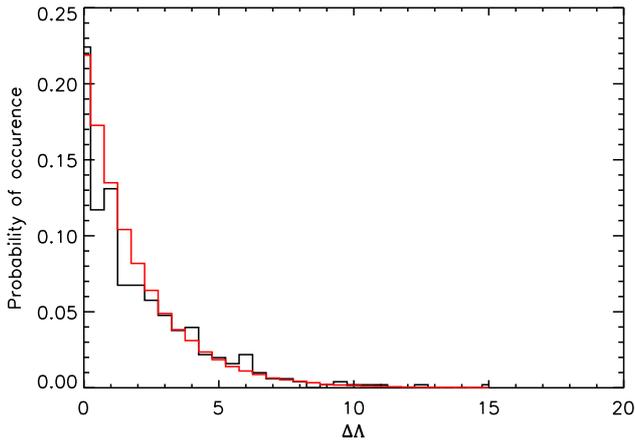}
\end{center}
\caption{Distribution of $\Delta\Lambda$ obtained with a Monte Carlo simulation with 500 iterations (black curve). The distribution of a $\chi^2$ with 2 degrees of freedom is overplotted (red curve).
\label{fig_simu_wilks}}
\end{figure}

To verify that $\Delta\Lambda$ is distributed as a $\chi^2$ with 2 degrees of freedom under the $H_0$ hypothesis, we performed a Monte Carlo simulation. We generated a mode profile with frequency, height, and linewidth typical of oscillation modes of stars in the clump. For each iteration, we included noise following the distribution of a $\chi^2$ with 2 degrees of freedom to simulate an observed power spectrum. Two fits were performed, either under the $H_0$ hypothesis (best likelihood $\ell_0$) or under the $H_1$ hypothesis (best likelihood $\ell_1$), and we stored the value of $\Delta\Lambda=-2\left[\ln(\ell1) - \ln(\ell_0)\right]$ for each iteration. Fig. \ref{fig_simu_wilks} shows the distribution that we obtained for $\Delta\Lambda$. For comparison, the distribution of a $\chi^2$ with 2 degrees of freedom is overplotted (red). We confirm that the two distributions are indeed very similar, which validates the assertion of \cite{wilks38} in our particular case.

\section{Modification of the \cite{goupil13} formula \label{app_G13}} 

We start by briefly recalling the method proposed by G13. By using approximate expressions of the eigenfunctions in the p-mode and g-mode cavities through asymptotic analysis, they approximated the mode trapping parameter $\zeta$ by the quantity
\begin{linenomath*}
\begin{equation}
\tilde{\zeta} = \left[ 1 + \left( \frac{c}{a} \right)^2 \frac{\theta_{\rm p}}{\theta_{\rm g}} \right]^{-1}
\label{eq_zetatilde_g13}
\end{equation}
\end{linenomath*}
where $\theta_{\rm p}$ and $\theta_{\rm g}$ were defined in Eq. \ref{eq_thetapg}. The coefficients $a$ and $c$ are the amplitudes of the eigenfunctions in the g-mode and p-mode cavities, respectively. The matching of the solutions in the evanescent zone provides the following relation between $a$ and $c$ (see \citealt{unno89}):
\begin{linenomath*}
\begin{equation}
\frac{c}{a} = 2 \frac{\cos(\theta_{\rm g})}{\cos(\theta_{\rm p})} \exp\left(\theta_{\rm e}\right)
\label{eq_csura}
\end{equation}
\end{linenomath*}
where
\begin{linenomath*}
\begin{equation}
\theta_{\rm e} \equiv \int_{r_{\rm b}}^{r_{\rm c}} \kappa \,dr
\end{equation}
\end{linenomath*}
and $\kappa^2=-k_r^2$ in the evanescent zone. We note that the factor $2 \exp(\theta_{\rm e})$ corresponds to $q^{-1/2}$, where the parameter $q$ was introduced in Eq. \ref{eq_matching} and whose value was already estimated from the mode frequencies for all the stars of the sample (see Sect. \ref{sect_deltapi}). We here made minor modifications to the expression of $\tilde{\zeta}$ proposed by G13 by releasing three approximations that they made. 
\begin{itemize}
\item First, in their equivalent of Eq. \ref{eq_csura}, G13 argued that $\cos{\theta_{\rm p}}\approx 1$ since $\theta_{\rm p}\approx n_{\rm p}\pi$. Technically, this is true only when the modes are p-dominated and should therefore not be assumed for all modes. We kept this term in the expression of $\tilde{\zeta}$.
\item Secondly, the approximate expressions for $\theta_{\rm p}$ and $\theta_{\rm g}$ used by G13 (Eq. A12 and A18 of their paper) in the phases of the cosine functions in Eq. \ref{eq_csura} neglect the phase shifts $\varepsilon_{\rm p}$ and $\varepsilon_{\rm g}$. Since we have included them to obtain estimates of $\Delta\nu$, $\Delta\pi_1$ and $q$ in Sect. \ref{sect_deltapi}, we used Eq. \ref{eq_thetap_m12} and \ref{eq_thetag_m12} for $\theta_{\rm p}$ and $\theta_{\rm g}$ respectively in Eq. \ref{eq_csura}, in order to be consistent.
\item Finally, G13 assumed that the factor $\exp(\theta_{\rm e})$ in Eq. \ref{eq_csura} was close to 1 owing to the narrow evanescent zones in red giants. As mentioned above, this factor corresponds to $(4q)^{-1/2}$, where $q$ has been measured from the mode frequencies. We found that this factor is indeed not far from unity (it varies from 0.7 to 1.1 for the stars of the sample) but taking it into account slightly improves the agreement with the actual values of $\zeta$.
\end{itemize}

In the end, we obtained the following modified expression for $\tilde{\zeta}$
\begin{linenomath*}
\begin{equation}
\tilde{\zeta} = \left \{ 1 + \frac{1}{q}\frac{\cos^2 \left[\pi\left(\frac{1}{\nu\Delta\Pi_1}-\varepsilon_{\rm g}\right)\right]}{\cos^2\left[\pi\frac{\left(\nu-\nu_{\rm p}\right)}{\Delta\nu}\right]} \frac{\nu^2\Delta\Pi_1}{\Delta\nu} \right \}^{-1}
\label{eq_zetatilde}
\end{equation}
\end{linenomath*}
This expression was used to estimate the mode trapping in a model of \cible\ in Sect. \ref{sect_G13} and an excellent agreement has been obtained with the $\zeta$ parameter computed with the mode eigenfunctions and Eq. \ref{eq_zeta} (see Fig. \ref{fig_compare_zeta}).

\end{appendix}

\end{document}